\documentclass[journal]{IEEEtran}
\usepackage{graphicx,amssymb,amsmath,color,textcomp,bm,cite,stfloats,algorithm,algorithmic,array,multirow,makecell,booktabs,subfigure,bbm,stmaryrd,amsmath,diagbox}
\usepackage[hidelinks]{hyperref}
\newtheorem{theorem}{Theorem}

\newcounter{MYtempeqncnt}

\begin{document}

\title{ 
Structured Tensor Decomposition Based Channel Estimation and Double Refinements for \\Active RIS Empowered Broadband Systems  
\thanks{ Yirun Wang, Yongqing Wang, and Yuyao Shen are with the School of Information and Electronics, Beijing
Institute of Technology, Beijing 100081, China   
(e-mail: yrwang719@163.com; wangyongqing@bit.edu.cn; syyxyz@gmail.com).  }
\thanks{ Gongpu Wang is with the Beijing Key Laboratory of Transportation Data Analysis and Mining, Beijing Jiaotong University, Beijing 100044,  China   
(e-mail: gpwang@bjtu.edu.cn).  }  
\thanks{ Chintha Tellambura is with the Department of Electrical and Computer
Engineering, University of Alberta,  Edmonton, AB T6G 2R3, Canada 
(e-mail: ct4@ualberta.ca). }
\thanks{Corresponding author: Yuyao Shen.}

\author{Yirun Wang, Yongqing Wang, Yuyao Shen, Gongpu Wang, and Chintha Tellambura, {\it  Fellow,  IEEE}}

}


\maketitle

\begin{abstract}

Channel parameter recovery is critical for the next-generation reconfigurable intelligent surface (RIS)-empowered communications and sensing. Tensor-based mechanisms are particularly effective, inherently capturing the multi-dimensional nature of wireless channels. However, existing studies assume either a line-of-sight (LOS) scenario or a blocked TX-RX channel. This paper solves a novel problem: tensor-based channel parameter estimation for active RIS-aided multiple-antenna broadband connections in fully multipath environments with the TX-RX link. System settings are customized to construct a fifth-order canonical polyadic (CP) signal tensor that matches the five-dimensional channel. Four tensor factors contain redundant columns, rendering the classical Kruskal's condition for decomposition uniqueness unsatisfied. The fifth-order Vandermonde structured CP decomposition (VSCPD) is developed to address this challenge, making the tensor factorization problem solvable using only linear algebra and offering a relaxed general uniqueness condition. With VSCPD as a perfect decoupling scheme, a sequential triple-stage channel estimation algorithm is proposed based on one-dimensional parameter estimation. The first stage enables multipath identification and algebraic coarse estimation. The following two stages offer optional successive refinements at the cost of increased complexity. The closed-form Cram\'er-Rao lower bound (CRLB) is derived to assess the estimation performance. Herein, the noise covariance matrix depends on multipath parameters in our active-RIS scenario. Numerical results are provided to verify the effectiveness of proposed algorithms under various evaluation metrics. Our results also show that active RIS can significantly improve channel estimation performance compared to passive RIS.


\end{abstract}


\section{Introduction} 

Reconfigurable intelligent surface (RIS) technology is being developed for B5G/6G and beyond. It can intelligently customize the radio propagation environment and enhance signal transmission coverage \cite{Wu2019a,Li2022d}. Generally, a RIS is a planar array composed of multiple reflection elements, each programmable to adjust the phase shift of incident signals \cite{Wu2021}. Optimizing the RIS reflection coefficients allows desired signals to be efficiently amplified \cite{Huang2019}, making RIS a promising solution to mitigate severe path loss in millimeter wave (mmWave) communications. Additionally, RIS can serve as a high-resolution reference point \cite{Liu2024b}, reducing reliance on conventional anchors in positioning systems.

RIS is categorized into several types, e.g., (nearly)-passive RIS \cite{Liu2021}, active RIS \cite{Long2021}, hybrid RIS \cite{Schroeder2022} (composed of reflectors and sensors), among others.   

As the RIS is a two-dimensional (2D) metasurface, the channel between it and a single-antenna radio can naturally be modeled as a matrix. The  RIS can be complemented with additional spatial dimensions by using multiple-antenna technology that enhances channel capacity and achieves diversity gain \cite{Stuber2004}. 
Furthermore, orthogonal frequency-division multiplexing (OFDM) offers an extra channel dimension -- the frequency domain, yielding superior robustness against multipath fading and improved spectrum efficiency \cite{Barhumi2003}. Therefore, when the RIS is integrated with multiple-antenna OFDM systems, the transmission channel can be inherently represented as a multi-dimensional array, i.e.,  tensor.

Accurate channel state information (CSI) is critical for the optimal design of RIS-empowered wireless networks. Among various techniques, tensor-based channel estimation (CE) has gained significant attention due to its ability to naturally represent multi-dimensional channels \cite{Chen2021}, decouple unknown parameters across dimensions \cite{Gong2023}, and achieve high estimation accuracy \cite{Zhou2016}.

CE is typically implemented using one of two approaches. The first directly reconstructs the channel coefficients from tensors or matrices \cite{Wei2021, Araujo2021, Du2023}. The second approach further models the channels using path-specific parameters such as path gain, delay, and angle-of-arrival/departure (AOA/AOD) \cite{Shi2022}, transforming the original CE problem into a parameter recovery issue. The challenge of the second approach lies in the high nonlinearity of the parameter estimation model. The efficient solution requires meticulous design, usually taking more operations than the first approach. Notably, acquiring channel parameters is crucial for subsequent sensing tasks like target localization \cite{Zhao2022a} and environment mapping \cite{Huang2023}. For instance, four different pseudoranges obtained from delays suffice to realize the perception of the three-dimensional (3D) location and clock offset. Therefore, this paper focuses on the second challenging approach, emphasizing parameter retrieval. 



Several works investigated tensor-based channel parameter estimation for RIS-aided wireless systems. 
Study \cite{Lin2021a} considered hybrid-RIS and developed an algebraic solution without iterations. 
However, a hybrid RIS setup requires an extra controller for RIS mode switchover and inter-equipment cooperation, increasing the implementation complexity. 
The work in \cite{Xu2022} proposed to turn off partial RIS elements to save pilot overhead and capitalized on extrapolation to complete the channel estimate. 
Reference \cite{Mo2023} studied the direct recovery of physical parameters from the signal tensor. 
In addition, references  \cite{Lin2022,Zheng2022,He2024} investigated the Vandermonde structured canonical polyadic decomposition (VSCPD)-based CE mechanisms. 
The VSCPD leverages the nature of Vandermonde factor matrices, offering accurate decomposition results with much lower complexity compared to conventional alternating least squares (ALS)-CPD algorithm \cite{Sorensen2013}. 
Reference \cite{Lin2022} contributed a twin-RIS structure and constructed a signal tensor with rank proportional to the RIS element number, thus yielding significant training overhead. 
Study \cite{Zheng2022} addressed this issue and significantly reduced the tensor rank to the multipath number.  Reference \cite{He2024} further considered a multiple-antenna user and investigated the extended CE issue. 

However, existing VSCPD-related studies \cite{Zheng2022,He2024} only formed a third-order tensor for CE, despite the channel having more than three dimensions. Consequently, distinct parameters could not be completely decoupled via tensor factorization, and then a 2D parameter estimation problem was involved. Optimal yet exhaustive 2D search was employed, resulting in significant computation overhead. Iterative methods like alternating optimization approaches \cite{Zhang2024,Zhang2022a,Zhang2024a} can mitigate the complexity, but a reliable initial value is required to guarantee global optimality. Note that during the initialization process and each iteration, the one-dimensional (1D) search may also be needed without closed-form solutions \cite{Zhang2022a}. Since the 2D optimization problem is challenging and computationally expensive to solve optimally, as mentioned above, there is a strong need to investigate higher-order tensor modeling that matches the number of channel dimensions and to study the related VSCPD.


Further, as for the system model, all the related studies above  \cite{Wei2021,Araujo2021,Du2023,Lin2021a,Xu2022,Mo2023,Lin2022,Zheng2022,He2024} assumed that the direct channel between user equipment (UE) and base station (BS) was blocked due to unfavorable environmental conditions, and the UE-BS connection entirely depended on the RIS to create an additional cascaded UE-RIS-BS link. As such, these existing algorithms were tailored for the CE of the cascaded UE-RIS-BS channel, benefiting from the problem simplification introduced by such block assumption. Under such a framework, when the direct UE-BS channel exists in other scenarios, the CE solution involves more implementation steps, incurring either performance loss or efficiency degradation.  Specifically, most studies considered eliminating the cascaded channel first to acquire the estimate of the direct channel, based on which the cascaded one could be next recovered \cite{Wei2021a,He2020}. This strategy inevitably deteriorates the estimation accuracy of the latter due to error propagation. 
Further, orthogonal RIS profile was devised to mitigate the interference between the direct and cascaded links \cite{Shtaiwi2021}. However, this approach inherently doubles the minimum required number of pilot blocks.


To address this issue, \cite{Zheng2024} considered the existence of the direct UE-BS link and utilized sequential CPDs to jointly recover parameters of both direct and cascade channels. 
Herein, the active RIS was deployed, enabling simultaneous signal reflection and amplification. 
As the power loss of the cascaded channel is the product of the two separate one-hop ones, the signal traveling from the cascaded link is far weaker than that from the direct link \cite{Wu2021}. 
Furthermore, during the training phase, the RIS phase shifts are usually chaotic and hence the spatial diversity gain at RIS is insufficient. 
Fortunately, active RIS can improve the strength of the cascaded channel to strike a balance between these two channels and thus can remarkably improve the system performance over the passive RIS \cite{Mylonopoulos2022,Zhang2023}.  
Nevertheless, study \cite{Zheng2024} only focused on the CE issue in a degraded complete line-of-sight (LOS) scenario compared with the forgoing relevant works \cite{Lin2021a,Xu2022,Mo2023,Lin2022,Zheng2022} in multipath cases. 
In areas with rich multipath components like urban and indoor environments, the accuracy of the CE algorithm in \cite{Zheng2024} greatly deteriorates owing to mismatched model, and the derived CE performance bound therein is also inapplicable anymore.

\begin{figure}[t]
	\centering
	\hspace{4.0cm}
	\includegraphics[width=0.95\linewidth]{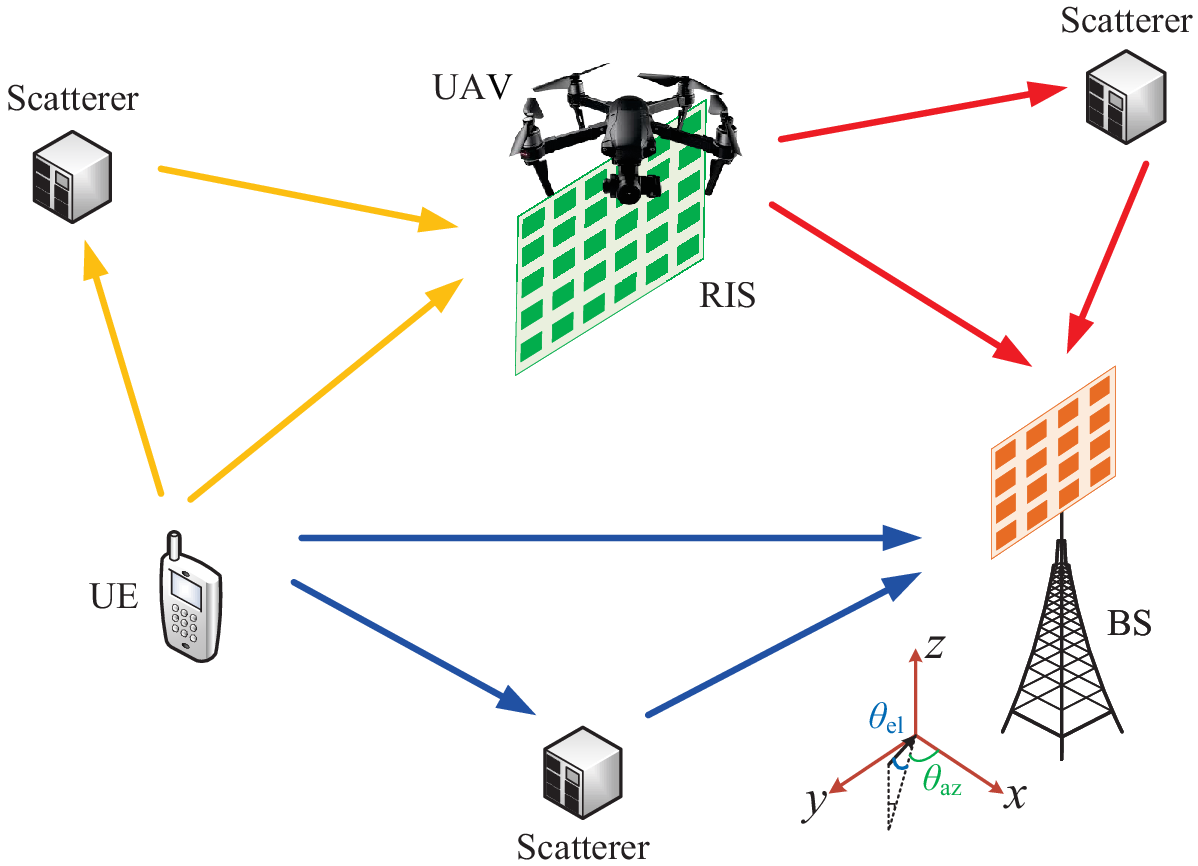}
	\caption{A UAV-mounted RIS-assisted uplink wireless system with multipath channels. 
	}
	\vspace {0.3cm}
	\label{fig:system}
\end{figure}

Motivated by the above, we extend the system model in \cite{Zheng2024} to the general multipath scenario and investigate the related tensor-based channel parameter acquisition issue for the first time. 
We formulate a CE scenario with active RIS deployed onto an unmanned aerial vehicle (UAV) to assist the uplink communications in a single-input multiple-output (SIMO)-OFDM system (Fig.~\ref{fig:system}). 
Both RIS and BS are assumed to form uniform planar array (UPA) topologies.
We consider the fully multipath case, i.e., frequency-selective channels for UE-BS, UE-RIS, and RIS-BS links. 
Owing to the UAV-mounted RIS's agile state, the CSI of RIS-BS link stays unavailable. 
We manage to recover physical parameters for direct UE-BS and cascaded UE-RIS-BS links via a fifth-order signal tensor. 
Any pre-processing steps of the received signal \cite{Wei2021a,He2020,Shtaiwi2021} are excluded. 
The contributions of this work are listed as follows.  
\begin{itemize}
	
	\item \textit{Tensor Construction:} 
	We carefully design the transmission pilots, RIS profile, and combining matrix. 
	Consequently, the collected received signals are reformulated as a fifth-order canonical polyadic (CP) tensor to match the five-dimensional channel. 
	Each factor matrix incorporates the entire information of one type of unknown parameters. 
    The tensor rank is the path number sum of direct and cascaded links.
	With the tensor factorization as a perfect decoupling manner, we convert the original multi-dimensional optimization problem into several parallel 1D estimation issues, successfully avoiding the mutual-coupling 2D optimization in \cite{Zheng2022,He2024}.
	\item \textit{Tensor Decomposition:} 
	Among the CP factors, the first is a Vandermonde matrix, while the other four all possess repeated columns. This configuration renders the classical Kruskal's uniqueness condition unsatisfied. To address this challenge, spatial smoothing-based fifth-order VSCPD is developed to enable tensor factorization. The VSCPD offers algebraic solution with desirable accuracy. A relaxed general uniqueness condition is provided, resulting in a loose demand relevant only to frequency-domain spatial-smoothing parameters for our case. 
	\item \textit{Parameter Estimation:}
	A tri-stage CE algorithm is proposed using factor matrix estimates. 
	The first stage performs multipath identification and gives algebraic coarse solution. By the end of this stage, only linear algebra and simple classification suffice to recover parameters.  
	Each of the subsequent two stages is optional, refining the estimation accuracy sequentially via search or iteration, thus at the expense of complexity growth. 
	\item \textit{Performance Bounding:} 
	We derive the closed-form Cram\'er-Rao lower bound (CRLB). 
	Owing to the additional introduction of thermal noise at active RIS, the covariance matrix of total received noise is also a function of multipath parameters, which was ignored in \cite{Zheng2024}. 
	The CRLB analysis in the conventional passive-RIS scenario is a particular case of ours with the active RIS. 
\end{itemize}

%

\begin{table}[t]
	\centering
	\caption{Summary of Main Acronyms}
	\renewcommand\arraystretch{1.1} 
	\begin{tabular}{ll}
		\Xhline{2.5\arrayrulewidth}
		Acronym & Description \\
		\hline
		1/2/3D & One/Two/Three-dimensional \\
		ALS & Alternating least squares \\
		AOA/AOD & Angle-of-arrival/departure \\
		BS & Base station  \\
		CBS & Correlation-based search  \\
		CE & Channel estimation \\
		CPD & Canonical polyadic decomposition  \\
		CRLB & Cramér-Rao lower bound   \\
		CSI & Channel state information \\
		LOS & Line-of-sight \\
		LS & Least squares \\
		ML & Maximum likelihood \\
		OFDM & Orthogonal frequency-division multiplexing \\
		RIS & Reconfigurable intelligent surface  \\
		SIMO & Single-input multiple-output  \\
		UAV & Unmanned aerial vehicle  \\
		UE & User equipment  \\
		VSCPD & Vandermonde structured CPD  \\
		\Xhline{2.5\arrayrulewidth}
	\end{tabular}%
	\label{tab:acronyms}%
\end{table}%

The main acronyms are listed in Table~\ref{tab:acronyms}. 
The remainder of this paper is organized as follows. 
Section~\ref{sec:PF} presents the channel and signal models. 
Section~\ref{sec:CE} describes tensor construction, VSCPD with spatial smoothing, and our triple-stage CE algorithm, as guided in Fig.~\ref{fig:flow_chart}.  
Section~\ref{sec:PA} derives CRLB and analyzes algorithm complexity. 
Section~\ref{sec:NR} provides numerical results and Section~\ref{sec:conc} concludes this paper.

{\it Notations:} Scalars, vectors, matrices, tensors, and sets are denoted by 
$ a $, $ \mathbf{a} $, $ \mathbf{A} $, $ \bm{\mathcal{A}} $, and $ \mathcal{A} $, respectively. 
$ \mathbb{R} $ and $ \mathbb{C} $ denote the real and complex domains, respectively.
The transpose, complex conjugate, conjugate transpose, inverse, and pseudo-inverse are denoted by $ (\cdot)^T $, $ (\cdot)^* $, $ (\cdot)^H $, $ (\cdot)^{-1} $, and $ (\cdot)^{\dagger} $, respectively. 
Vectorization, reverse vectorization, and trace are denoted by 
$ \operatorname{vec}(\cdot) $, $ \operatorname{unvec}(\cdot) $, $ \operatorname{tr}(\cdot) $, respectively. 
Real part, imaginary part, and amplitude are denoted by $ \mathfrak{R}(\cdot) $, $ \mathfrak{I}(\cdot) $, and $ |\cdot| $ respectively, and $ \jmath^2=-1 $. 
$ \operatorname{diag}\{\mathbf{a}\} $ denotes the diagonal matrix formed by $ \mathbf{a} $.  
$ \underline{\mathbf{A}} $ is the submatrix deleting the last row of $ \mathbf{A} $. 
$ \|\cdot\| $, $ \|\cdot\|_F $, $ \llbracket \cdot \rrbracket $, and $ \mathbb{E}[\cdot] $ denote the Euclidean norm, Frobenius norm, CPD operator, and statistical expectation, respectively. 
Symbols $ \circ $, $ \otimes $, $ \odot $, and $ \circledast $ denote the outer product, Kronecker product, Khatri–Rao product, and Hadamard product, respectively. 
$ \mathbf{1} $, $ \mathbf{0} $, and $ \mathbf{I} $ denote the all-one, all-zero, and identity matrices, respectively. 
$ \operatorname{rank}(\mathbf{A}) $ and $ k_{\mathbf{A}} $ denote the matrix rank and Kruskal rank (or $k$-rank) of $ \mathbf{A} $, respectively. 
Finally, $ \mathcal{N}(\bm{\mu},\bm{\Sigma}) $ and $ \mathcal{CN}(\bm{\mu},\bm{\Sigma}) $ denote real and complex Gaussian distributions with mean $ \bm{\mu} $ and variance $ \bm{\Sigma} $, respectively. 

{\it Vandermonde matrix:} Matrix $ \mathbf{A}\in\mathbb{C}^{I\times R} $ is called a Vandermonde matrix if $ [\mathbf{A}]_{i,r}=e^{\jmath(i-1)\omega_r} $, and $ \{\omega_r\}_{r=1}^R $ are its generators.

\section{Problem Formulation} \label{sec:PF}

This work considers attaching a RIS to a UAV to assist uplink communications from a UE to the BS -- Fig.~\ref{fig:system}. The fully multipath scenario is assumed, including the presence of a direct UE-BS channel, to address the general CE issue. Additionally, the UE is assumed to be a single-antenna radio. Meanwhile, both BS and RIS employ UPA of sizes  $ \tilde{N}_y\times \tilde{N}_z $ and $ M_y \times M_z $, respectively, located on the YOZ plane in their local coordinate systems (LCSs).  
The differences between our UAV-mounted RIS case and fixed RIS one are presented at the end of this section. 

This paper resorts to the active RIS. Each element of it comprises a phase shifter and an active amplifier for incident signals. 
Active RIS strengthens the power of the cascaded UE-RIS-BS channel to achieve a balance between the cascaded and direct UE-BS channels. 
The amplification coefficients of  RIS elements are assumed to remain identical and denoted by unified notation $ \eta>1 $. 
It is worth noting that the signal amplification at the active RIS introduces additional thermal noise, which is neglected in the conventional passive-RIS case. 
The energy consumption of active RIS can be modeled as \cite{Peng2022}
\begin{equation}\label{eq:PR}
	P_{\textrm{R}} = 
	(\eta^2-1)  M_yM_z  (P_{\textrm{in}}+\sigma_{\textrm{R}}^2),
\end{equation}
where $ P_{\textrm{in}} $ is the average power of incident signals at each RIS element, 
and $ \sigma_{\textrm{R}}^2 $ denotes the power of the thermal noises generated from every RIS element. 

\subsection{Channel Model}

This paper considers an OFDM broadband wireless system over frequency-selective fading channels. The channel frequency responses (CFRs) for the UE-BS, UE-RIS, and RIS-BS links associated with the $k$th subcarrier are denoted by 
$ \mathbf{h}_{\textrm{L}}^{(k)} \in\mathbb{C}^{\tilde{N}_y\tilde{N}_z \times 1} $, 
$ \mathbf{h}_ {\textrm{R},1} ^{(k)} \in\mathbb{C}^{M_yM_z \times 1} $, and
$ \mathbf{H}_ {\textrm{R},2} ^{(k)} \in\mathbb{C}^{\tilde{N}_y\tilde{N}_z \times M_yM_z } $  respectively, 
$ k=1,\ldots,K_0 $ with total $ K_0 $ subcarriers. 
The channels are formulated using geometry-related  multipath parameters as 
\begin{align}
	\mathbf{h}_{\textrm{L}}^{(k)} &= 
	\sum_{\ell=1}^{L} 
	\beta_{\textrm{L}}^{(\ell)} e^{-\jmath 2 \pi f^{(k)} \tau_{\textrm{L}}^{(\ell)}} 
	\mathbf{a}_{\textrm{B}} \big( \bm{\theta}_{\textrm{L}}^{(\ell)} \big),
	\\ 
	\mathbf{h}_ {\textrm{R},1} ^{(k)} = & 
	\sum_{p=1}^{P}
	\beta_{\textrm{R},1}^{(p)} e^{-\jmath 2 \pi f^{(k)} \tau_{\textrm{R},1}^{(p)}} 
	\mathbf{a}_{\textrm{R}} \big( \bm{\varphi}_{\textrm{A}}^{(p)} \big),
	\label{eq:hR1}
	\\
	\mathbf{H}_ {\textrm{R},2} ^{(k)} = &
	\sum_{q=1}^{Q}
	\beta_{\textrm{R},2}^{(q)} e^{-\jmath 2 \pi f^{(k)} \tau_{\textrm{R},2}^{(q)}} 
	\mathbf{a}_{\textrm{B}} \big( \bm{\theta}_{\textrm{R}}^{(q)} \big)
	\mathbf{a}_{\textrm{R}}^T \big( \bm{\varphi}_{\textrm{D}}^{(q)} \big),
	\label{eq:HR2}
\end{align}
where $ L $, $ P $, and $ Q $ denote the number of paths for UE-BS, UE-RIS, and RIS-BS channels, respectively, 
$ \beta_{\textrm{L}}^{(\ell)} $, $ \beta_{\textrm{R},1}^{(p)} $, and $ \beta_{\textrm{R},2}^{(q)} $ represent the path gains, 
$  \tau_{\textrm{L}}^{(\ell)} $, $ \tau_{\textrm{R},1}^{(p)} $, and $ \tau_{\textrm{R},2}^{(q)} $ denote channel delays, 
$ f^{(k)} = (k-1)\Delta_f $ is subcarrier frequency with $ \Delta_f $ being frequency spacing, 
$ \mathbf{a}_{\textrm{B}}\in\mathbb{C}^{\tilde{N}_y\tilde{N}_z \times 1} $ and $ \mathbf{a}_{\textrm{R}}\in\mathbb{C}^{M_yM_z \times 1} $ are the array responses of BS and RIS, respectively,  
$ \bm{\theta}_{\textrm{L}}^{(\ell)}=
\big[ \theta_{\textrm{L},\textrm{az}}^{(\ell)}, \theta_{\textrm{L},\textrm{el}}^{(\ell)} \big]^T $ and 
$ \bm{\theta}_{\textrm{R}}^{(\ell)}=
\big[ \theta_{\textrm{R},\textrm{az}}^{(\ell)}, \theta_{\textrm{R},\textrm{el}}^{(\ell)} \big]^T $ 
denote the AOA pairs of BS for UE-BS and RIS-BS channels, respectively, with each pair containing an azimuth angle and an elevation angle, 
similarly 
$ \bm{\varphi}_{\textrm{A}}^{(p)} =
\big[ \varphi_{\textrm{A},\textrm{az}}^{(p)}, \varphi_{\textrm{A},\textrm{el}}^{(p)} \big]^T $ and 
$ \bm{\varphi}_{\textrm{D}}^{(q)} = 
\big[ \varphi_{\textrm{D},\textrm{az}}^{(q)}, \varphi_{\textrm{D},\textrm{el}}^{(q)} \big]^T $ 
represent RIS's AOA and AOD  pairs for the UE-RIS and RIS-BS channels.

The spatial array response is formulated as follows. Take the RIS's one as a detailed example. 
Without loss of generality, assume the RIS elements start from the local coordinate origin. 
Out of unified expression, temporarily remove some subscripts and superscripts for specific individuals, and $ \mathbf{a}_{\textrm{R}} (\bm{\varphi}) $ can be formed from 
\begin{equation}\label{eq:a_org}
	[\mathbf{a}_{\textrm{R}} (\bm{\varphi})]_m = 
	e^{\jmath \frac{2\pi}{\lambda} [\mathbf{P}_{\textrm{R}}]_{m,:} \mathbf{d}(\bm{\varphi})},
	\; m=1,\ldots,M_yM_z, 
\end{equation}
where 
$ \mathbf{P}_{\textrm{R}} = 
\big[ \mathbf{0}_{M_yM_z\times1}, 
\mathbf{y}_{\textrm{R}} \otimes \mathbf{1}_{M_z\times1}, 
\mathbf{1}_{M_y\times1} \otimes\mathbf{z}_{\textrm{R}} \big] 
\in\mathbb{R}^{M_yM_z\times3} $ 
vertically stacks the local 3D positions of every RIS element, 
$ \mathbf{y}_{\textrm{R}}\in\mathbb{R}^{M_y\times1} $ and $ \mathbf{z}_{\textrm{R}}\in\mathbb{R}^{M_z\times1} $ denote the local Y positions and Z positions of RIS elements, respectively, 
$ \lambda $ is the signal wavelength, and 
$ \mathbf{d}(\bm{\varphi}) = 
\big[ \cos(\varphi_{\textrm{az}}) \cos(\varphi_{\textrm{el}}),
 \sin(\varphi_{\textrm{az}}) \cos(\varphi_{\textrm{el}}),
 \sin(\varphi_{\textrm{el}}) \big]^T $  
represents the direction vector of the arrival/departure signal. 
Further, assume the RIS array is uniformly spaced on both two axes 
with identical element spacing $ d_{\textrm{R}} $, 
then \eqref{eq:a_org} can be recast as
\begin{equation}\label{eq:aR}
	\mathbf{a}_{\textrm{R}} (\bm{\varphi}) = 
	\mathbf{a}^{(M_y)} (\omega_{2}) \otimes 
	\mathbf{a}^{(M_z)} (\omega_{3}), 
\end{equation}
where $ \big[\mathbf{a}^{(M)} (\omega)\big]_m = 
e^{ \jmath (m-1)\omega }, m=1,\ldots,M $, and 
\begin{equation}
	\omega_{2} = \frac{2\pi}{\lambda} d_{\textrm{R}} \sin(\varphi_{\textrm{az}}) \cos(\varphi_{\textrm{el}}), \;
	\omega_{3} = \frac{2\pi}{\lambda} d_{\textrm{R}} \sin(\varphi_{\textrm{el}}).
\end{equation} 
Analogously, the array response $ \mathbf{a}_{\textrm{B}} (\bm{\theta}) $ of BS can be expressed as
\begin{gather}
	\label{eq:aB}
	\mathbf{a}_{\textrm{B}} (\bm{\theta}) = 
	\mathbf{a}^{(\tilde{N}_y)} (\omega_{4}) \otimes 
	\mathbf{a}^{(\tilde{N}_z)} (\omega_{5}),  \\
	\omega_{4} = \frac{2\pi}{\lambda} d_{\textrm{B}} \sin(\theta_{\textrm{az}}) \cos(\theta_{\textrm{el}}), \;
	\omega_{5} = \frac{2\pi}{\lambda} d_{\textrm{B}} \sin(\theta_{\textrm{el}}),
\end{gather}
and  
we denote the BS's position matrix and element spacing by 
$ \mathbf{P}_{\textrm{B}} $ and $ d_{\textrm{B}} $, respectively.

\subsection{Signal Model}\label{sec:SM}

Consider that the UE transmits $ G $ OFDM symbols to the BS in the time domain. 
We assume that the active-RIS profile 
$ \bm{\gamma}^{(g)} = \eta \big[  e^{\jmath\vartheta_1^{(g)}}, \ldots, 
e^{\jmath\vartheta_{M_yM_z}^{(g)}} \big]^T
\in\mathbb{C}^{M_yM_z\times1}  $ 
changes at every distinct time slot $ g $ by adjusting phase shifts, $ g=1,\ldots,G $. 
Let $ \mathbf{\Gamma}^{(g)}=\operatorname{diag}(\bm{\gamma}^{(g)}) $. 
The received baseband signal for the $ k $th subcarrier and the $ g $th slot is given by 
\begin{equation}
	\tilde{\mathbf{y}}^{(k,g)} =
	\mathbf{R}^H \big(  \mathbf{h}_{\textrm{L}}^{(k)}  + 
	\mathbf{h}_{\textrm{R}}^{(k,g)}  \big) x^{(k,g)} + 
	\mathbf{w}^{(k,g)},
	\label{eq:ykg2}
\end{equation}
where
$ \mathbf{R} \in\mathbb{C}^{ \tilde{N}_y\tilde{N}_z \times N_1N_2  } $ denotes the combining matrix, 
$ x^{(k,g)} $ is the transmitted signal with transmit power $ |x^{(k,g)}|^2 = P_{\textrm{T}} $, 
$ \mathbf{h}_{\textrm{R}}^{(k,g)} = 
\mathbf{H}_ {\textrm{R},2} ^{(k)}  \mathbf{\Gamma}^{(g)}  \mathbf{h}_ {\textrm{R},1} ^{(k)} $ 
denotes the cascaded UE-RIS-BS channel, 
and the overall received noise $ \mathbf{w} $ is expressed as 
\begin{equation}\label{eq:w}
	\mathbf{w}^{(k,g)} = 
	\mathbf{R}^H
	\big( \mathbf{w}_{\textrm{B}}^{(k,g)} +  \mathbf{H}_ {\textrm{R},2} ^{(k)}  \mathbf{\Gamma}^{(g)}  \mathbf{w}_{\textrm{R}}^{(k,g)}  
	 \big),
\end{equation}
where 
$ \mathbf{w}_{\textrm{B}}^{(k,g)} \sim 
\mathcal{CN}(\mathbf{0}_{\tilde{N}_y\tilde{N}_z\times1}, \sigma_{\textrm{B}}^2\mathbf{I}_{\tilde{N}_y\tilde{N}_z}) $ 
denotes the received thermal noise at BS with noise power $ \sigma_{\textrm{B}}^2 $, and 
$ \mathbf{w}_{\textrm{R}}^{(k,g)} \sim 
\mathcal{CN}(\mathbf{0}_{M_yM_z\times1}, \sigma_{\textrm{R}}^2\mathbf{I}_{M_yM_z}) $ 
is the thermal noise incurred by active RIS elements.
Noises $ \mathbf{w}_{\textrm{B}}^{(k,g)} $ and $ \mathbf{w}_{\textrm{R}}^{(k,g)} $ are uncorrelated to each other. 
Notably, $ \mathbf{w}^{(k,g)} $ depends on the unknown RIS-BS channel 
$  \mathbf{H}_ {\textrm{R},2}^{(k)} $ here in our active-RIS scenario, and \eqref{eq:w} degrades into the conventional passive-RIS case if removing the second noise term. 
These noise results are important for the bounding issue in Section~\ref{sec:CRLB}. 

Introducing \eqref{eq:hR1}\textendash\eqref{eq:HR2}, cascaded channel $ \mathbf{h}_{\textrm{R}}^{(k,g)} $ in \eqref{eq:ykg2} further becomes
\begin{equation}
	\mathbf{h}_ {\textrm{R}} ^{(k,g)}  = 
	\sum_{p=1}^{P} \sum_{q=1}^{Q} 
	\beta_{\textrm{R}}^{(p,q)} \rho_{\textrm{R}}^{(p,q)}
	e^{-\jmath 2 \pi f^{(k)} \tau_{\textrm{R}}^{(p,q)}}
	\mathbf{a}_{\textrm{B}} \big( \bm{\theta}_{\textrm{R}}^{(q)} \big),
\end{equation}
where 
$ \beta_{\textrm{R}}^{(p,q)} = \beta_{\textrm{R},1}^{(p)}\beta_{\textrm{R},2}^{(q)} $ 
denotes the cascaded path gain, 
$ \tau_{\textrm{R}}^{(p,q)} = \tau_{\textrm{R},1}^{(p)} + \tau_{\textrm{R},2}^{(q)} $ 
represents the cascaded delay, 
and 
\begin{equation}
	\rho_{\textrm{R}}^{(p,q)} =
	\big(\bm{\gamma}^{(g)}\big)^T
	\Big(  \mathbf{a}_{\textrm{R}} \big( \bm{\varphi}_{\textrm{A}}^{(p)} \big)
	\circledast
	\big( \mathbf{a}_{\textrm{R}} \big( \bm{\varphi}_{\textrm{D}}^{(q)} \big) \Big). 
	\label{eq:rho0}
\end{equation}
Recall \eqref{eq:aR} and therein we restore individual subscripts and superscripts, then \eqref{eq:rho0} can be rewritten as \cite[Eq.~(4)]{Huang2021}
\begin{equation}
	\rho_{\textrm{R}}^{(p,q)} = 
	\big(\bm{\gamma}^{(g)}\big)^T 
	\left(
	\mathbf{a}^{(M_y)} (\omega_{2}^{(p,q)})  \otimes
	\mathbf{a}^{(M_z)} (\omega_{3}^{(p,q)}) 
	\right), 
\end{equation}
where 
$ \omega_{2}^{(p,q)} = \omega_{\textrm{A},2}^{(p)} + \omega_{\textrm{D},2}^{(q)} = 
\frac{2\pi}{\lambda} d_{\textrm{R}} \psi_{2}^{(p,q)}  $ and 
$ \omega_{3}^{(p,q)} = \omega_{\textrm{A},3}^{(p)} + \omega_{\textrm{D},3}^{(q)} = 
\frac{2\pi}{\lambda} d_{\textrm{R}} \psi_{3}^{(p,q)}  $ 
denote the effective RIS-incurred phase-shift spacings along the Y and Z axes, respectively, with intermediate angle-related parameters defined as
\begin{align}
	\psi_{2}^{(p,q)}
	&= \sin (\varphi_{\textrm{A},\textrm{az}}^{(p)})  \cos(\varphi_{\textrm{A},\textrm{el}}^{(p)})+
	\sin(\varphi_{\textrm{D},\textrm{az}}^{(q)}) \cos(\varphi_{\textrm{D},\textrm{el}}^{(q)}),
	\nonumber \\
	\psi_{3}^{(p,q)}
	&= \sin (\varphi_{\textrm{A},\textrm{el}}^{(p)})+
	\sin (\varphi_{\textrm{D},\textrm{el}}^{(q)}).
\end{align}

Note that in contrast to the direct link, the path gain of the cascaded link is far weaker owing to its concatenation nature, i.e., 
$ \big|\beta_{\textrm{R}}^{(p,q)}\big| = \big|\beta_{\textrm{R},1}^{(p)}\big|\big|\beta_{\textrm{R},2}^{(q)}\big| \ll \big|\beta_{\textrm{L}}^{(\ell)}\big|, \forall p,q,\ell $. 
Moreover, we cannot fully reap the RIS beamforming gain during the training phase owing to the unavailable CSI. 
To improve the power of the cascaded channel, we resort to the active RIS enabling signal amplification. 
A comparison between passive-RIS and active-RIS systems can be found in Section \ref{sec:2modes_com}.

This paper assumes that $ K $ training subcarriers are provided among all $ K_0 $ subcarriers. Without loss of generality, we further assume that the first $ K $ subcarriers (i.e., $ k=1,\ldots,K $) are designed for pilot transmission for ease of exposition. The extension to the general case with equispaced training subcarriers is straightforward. 
We aim to extract unknown multipath parameters for direct UE-BS and cascaded UE-RIS-BS links to recover these two channels using received training signals. The parameters of interest are: 
delays $ \{ \tau_{\textrm{L}}^{(\ell)} \} \cup \{\tau_{\textrm{R}}^{(p,q)}\}  $, 
angle-related parameters at RIS $ \{ \psi_{2}^{(p,q)},\psi_{3}^{(p,q)} \} $, 
AOAs at BS  $ \{ \bm{\theta}_{\textrm{L}}^{(\ell)} \} \cup \{\bm{\theta}_{\textrm{R}}^{(q)}\}  $, and 
path gains $ \{ \beta_{\textrm{L}}^{(\ell)} \} \cup \{\beta_{\textrm{R}}^{(p,q)}\} $.\footnote{This paper only considers the estimation of cascaded UE-RIS-BS channel, which suffices to meet the demand of most subsequent tasks such as location awareness \cite{Zheng2024} and beamforming design \cite{Peng2023}. Without extra information, the acquisition of separate one-hop channels for UE-RIS and RIS-BS links is infeasible due to inherent ambiguities. For example, the cascaded path gain satisfies that $ \beta_{\textrm{R}}^{(p,q)} = \beta_{\textrm{R},1}^{(p)}\beta_{\textrm{R},2}^{(q)} =  \big(\zeta\beta_{\textrm{R},1}^{(p)}\big) \big(\frac{1}{\zeta}\beta_{\textrm{R},2}^{(q)}\big) $ with arbitrary scaling ambiguity parameter $ \zeta $. Nonetheless, the separate channel estimation can be easily achieved by employing hybrid RIS \cite{Lin2021a}, which is beyond the context of this current work.} Note that $ \{ \psi_{2}^{(p,q)},\psi_{3}^{(p,q)} \} $ are generated from the cascaded UE-RIS-BS channel and not related to the direct UE-BS link. 
In addition, $ \{\bm{\theta}_{\textrm{R}}^{(q)}\} $ merely come from the RIS-BS channel, irrelevant to the UE-RIS one. 
These two findings will be leveraged in the following algorithm development presented in the next section. 

Some discussions on our UAV-mounted RIS scenario can be found in the sequel. 
In traditional cases with fixed RIS deployed on walls, roadsides, etc., the CSI for RIS-BS link can be accurately measured in advance, serving as prior information to simplify the estimation of other channels \cite{Wan2020}.  
Nevertheless, this paper discusses a general CE case without any RIS-related information known \textit{a priori} as the result of the agile state of UAV-mounted RIS. 
Reversely, with the recovery of channel parameters, we can further determine the RIS state (e.g., location and orientation) \cite{Lu2022}. 
From the system design aspect, the customizable state of UAV-mounted RIS also provides additional degree of freedoms (DOFs) in optimizing system performances \cite{Li2020}. 

\begin{figure}[t]
	\centering
	\includegraphics[width=1\linewidth]{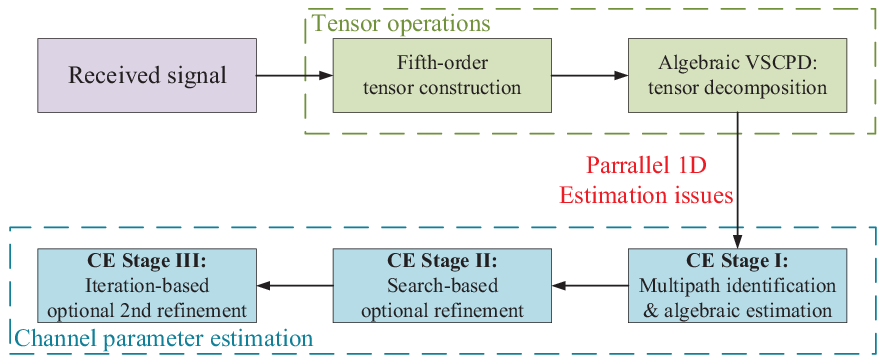}
	\caption{A flowchart of the proposed tensor-based CE framework.}
	\label{fig:flow_chart}
\end{figure}

\section{Proposed Tensor-Based CE Framework}\label{sec:CE}

We outline a flowchart of our tensor-based CE framework in Fig.~\ref{fig:flow_chart}. 
This section will present the step-by-step process in great detail. 
Our CE framework is developed under a noise-free model to illuminate its rationale. 
In real-world cases, the noise-free signal should be replaced by noisy observation as the algorithm input.

\subsection{Tensor Construction}
 
We define from \eqref{eq:ykg2} that 
$ \mathbf{y}^{(k,g)}_{\textrm{L}} = \mathbf{R}^H \mathbf{h}_{\textrm{L}}^{(k)} x^{(k,g)} 
\in\mathbb{C}^{N_1 N_2\times1} $ and 
$ \mathbf{y}^{(k,g)}_{\textrm{R}} = \mathbf{R}^H \mathbf{h}_{\textrm{R}}^{(k,g)} x^{(k,g)} $ 
represent the noise-free received signals traveled from direct and cascaded channels, respectively. 
Vertically stack 
$ \mathbf{y}_{\textrm{L}}^{(k,g)} $ and $ \mathbf{y}_{\textrm{R}}^{(k,g)} $
over time slot $ g $'s, respectively, yielding 
\begin{align}\setcounter{equation}{15}
	\mathbf{y}_{\textrm{L}}^{(k)} &= 
	\sum_{\ell=1}^{L} 
	\beta_{\textrm{L}}^{(\ell)} e^{-\jmath 2 \pi f^{(k)} \tau_{\textrm{L}}^{(\ell)}} 
	\big( \mathbf{x}^{(k)} \big) \otimes 
	\big(\mathbf{R}^H \mathbf{a}_{\textrm{B}} \big( \bm{\theta}_{\textrm{L}}^{(\ell)} \big)\big),
	\label{eq:yLk}
	\\
	\mathbf{y}_{\textrm{R}}^{(k)} & = 
	\sum_{p=1}^{P} \sum_{q=1}^{Q} 
	\beta_{\textrm{R}}^{(p,q)} e^{-\jmath 2 \pi f^{(k)} \tau_{\textrm{R}}^{(p,q)}} 
	\Big( \operatorname{diag}(\mathbf{x}^{(k)}) \mathbf{\Upsilon}  \nonumber \\
	&\times \big( \mathbf{a}^{(M_y)} (\omega_{2}^{(p,q)})  \otimes
	\mathbf{a}^{(M_z)} (\omega_{3}^{(p,q)}) \big) \Big) \otimes
	\big( \mathbf{R}^H\mathbf{a}_{\textrm{B}} \big( \bm{\theta}_{\textrm{R}}^{(q)} \big) \big),&
	\nonumber \label{eq:yRk}
\end{align}
where 
$ \mathbf{y}_{\textrm{L(R)}}^{(k)} = \big[ 
\mathbf{y}_{\textrm{L(R)}}^{(k,1)}, \ldots, \mathbf{y}_{\textrm{L(R)}}^{(k,G)} \big]^T 
\in\mathbb{C}^{GN_1N_2\times1} $, 
$ \mathbf{x}^{(k)} = [x^{(k,1)},\ldots,x^{(k,G)}]^T \in\mathbb{C}^{G\times1} $ denotes the pilot collection for the $ k $th subcarrier, and 
$ \mathbf{\Upsilon} = [\bm{\gamma}^{(1)},\ldots,\bm{\gamma}^{(G)}]^T \in\mathbb{C}^{G\times M_yM_z } $ is the RIS profile matrix over all time slots.  

We customize the transmission pilots, RIS profile, the combining matrix to facilitate channel estimation. 
Firstly,  identical pilots are selected  across time slots and training subcarriers, i.e., 
$ x^{(k,g)} = x^{(k',g')},\, k,k'\in\{1,\ldots,K\},\;g, g'\in\{1,\ldots,G\} $, and thus omit the superscripts of pilot notations hereafter. 
In this case, it is true that 
\begin{equation}\label{eq:x}
	\mathbf{x} = x\mathbf{1}_{G\times1} = (\sqrt{x}\mathbf{1}_{G_1\times1}) \otimes (\sqrt{x}\mathbf{1}_{G_2\times1})  
\end{equation}
with $ G = G_1 G_2 $. 
Secondly, we design
\begin{equation}
	\operatorname{diag}(\mathbf{x}) \mathbf{\Upsilon} = 
	x \mathbf{\Upsilon} = 
	\mathbf{T}_2^H \otimes \mathbf{T}_3^H  
\end{equation}
with carefully preset Vandermonde matrices 
$ \mathbf{T}_2\in\mathbb{C}^{M_y\times G_1} $ and 
$ \mathbf{T}_3\in\mathbb{C}^{M_z\times G_2} $. 
The solution of RIS profile matrix $ \mathbf{\Upsilon} $ lies in 
$ \mathbf{\Upsilon} = 
\frac{1}{x} \mathbf{T}_2^H \otimes \mathbf{T}_3^H $. 
Thirdly,  the combining matrix is set as 
\begin{equation}
	\mathbf{R} = \mathbf{T}_4 \otimes \mathbf{T}_5, 
\end{equation}
where $ \mathbf{T}_4\in\mathbb{C}^{\tilde{N}_y\times N_1} $ and 
$ \mathbf{T}_5\in\mathbb{C}^{\tilde{N}_z\times N_2} $ are also preconstructed Vandermonde matrices. 
The special structure of $ \{\mathbf{T}_n\}_{n=2}^5 $ will be utilized in Sections~\ref{sec:ESPRIT}~and~\ref{sec:CEI}. 

With the designs above, further vertically collecting $ \mathbf{y}_{\textrm{L}}^{(k)} $ and  $ \mathbf{y}_{\textrm{R}}^{(k)} $ in \eqref{eq:yLk} across training subcarrier $ k $'s, respectively, it follows that  
\begin{align}
	\label{eq:yLR}
	&\mathbf{y}_{\textrm{L}} = 
	\sum_{\ell=1}^{L} 
	\beta_{\textrm{L}}^{(\ell)} 
	\big( \mathbf{a}^{(K)} (\omega^{(\ell)}_{1,\textrm{L}}) \big)  \otimes
	(\sqrt{x}\mathbf{1}_{G_1\times1})   \\
	& \otimes (\sqrt{x}\mathbf{1}_{G_2\times1})
	 \otimes \big( \mathbf{T}_4^H \mathbf{a}^{(\tilde{N}_y)} (\omega_{4,\textrm{L}}^{(\ell)}) \big) \otimes 
	\big( \mathbf{T}_5^H \mathbf{a}^{(\tilde{N}_z)} (\omega_{5,\textrm{L}}^{(\ell)}) \big),&
	\nonumber \\
	&\mathbf{y}_{\textrm{R}}  = 
	\sum_{p=1}^{P} \sum_{q=1}^{Q} 
	\beta_{\textrm{R}}^{(p,q)} 
	\big( \mathbf{a}^{(K)} (\omega^{(p,q)}_{1,\textrm{R}}) \big) \otimes 
	\big( \mathbf{T}_2^H \mathbf{a}^{(M_y)} (\omega_{2}^{(p,q)}) \big) \nonumber \\ 
	&\!\!\otimes\!	\big( \mathbf{T}_3^H \mathbf{a}^{(M_z)} (\omega_{3}^{(p,q)}) \big)\!
	\otimes\! \big( \mathbf{T}_4^H \mathbf{a}^{(\tilde{N}_y)} (\omega_{4,\textrm{R}}^{(q)}) \big)\! \otimes\! 
	\big( \mathbf{T}_5^H \mathbf{a}^{(\tilde{N}_z)} (\omega_{5,\textrm{R}}^{(q)}) \big), &
	\nonumber
\end{align}
where we have capitalized on \eqref{eq:aB} and added individual subscripts and superscripts, 
$ \mathbf{y}_{\textrm{L}},\mathbf{y}_{\textrm{R}} \in\mathbb{C}^{KG_1G_2N_1N_2\times1} $, and 
\begin{equation}
	\omega^{(\ell)}_{1,\textrm{L}} =
	-2\pi\Delta_f \tau_{\textrm{L}}^{(\ell)}, \;
	\omega^{(p,q)}_{1,\textrm{R}} = 
	-2\pi\Delta_f \tau_{\textrm{R}}^{(p,q)}.
\end{equation}

\begin{figure}[t]
	\centering
	\includegraphics[width=1\linewidth]{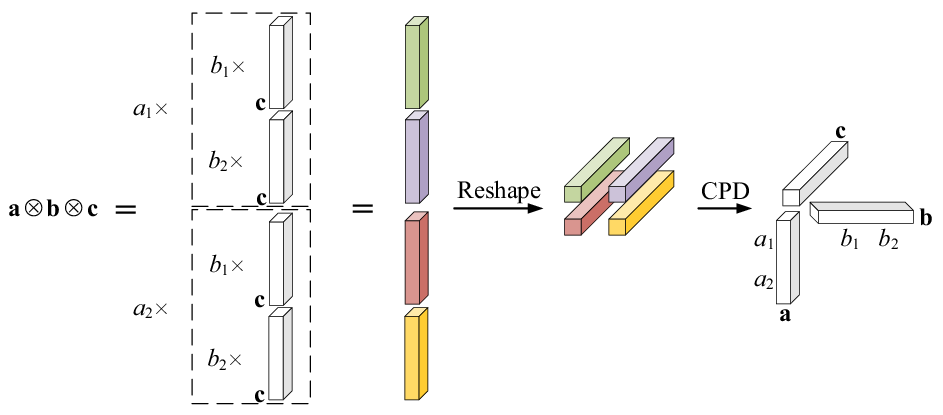}
	\caption{Illustration of reshaping Kronecker product of vectors into a tensor. We consider an example with $ \mathbf{a}\otimes\mathbf{b}\otimes\mathbf{c} $ resulting in a third-order rank-$1$ tensor, where both $ \mathbf{a} $ and $ \mathbf{b} $ only contain two elements for ease of illustration. 	}
	\label{fig:tensor}
\end{figure}

Reshape 
$ \mathbf{y}_{\textrm{L}} $ and $ \mathbf{y}_{\textrm{R}} $ in \eqref{eq:yLR} into corresponding 
tensors 
$ \bm{\mathcal{Y}}_{\textrm{L}}, \bm{\mathcal{Y}}_{\textrm{R}} 
\in\mathbb{C}^{K\times G_1\times G_2\times N_1\times N_2} $ and let 
$ \bm{\mathcal{Y}} = \bm{\mathcal{Y}}_{\textrm{L}} + \bm{\mathcal{Y}}_{\textrm{R}} $, yielding 
\begin{equation}\label{eq:tensor}
	\bm{\mathcal{Y}} = \llbracket \bm{\beta};
	\mathbf{A}_1, \mathbf{B}_2, \mathbf{B}_3, \mathbf{B}_4, \mathbf{B}_5 \rrbracket, 
\end{equation}
where 
the weight vector $ \bm{\beta} $ and factor matrices $ \mathbf{A}_1,\mathbf{B}_2,\ldots,\mathbf{B}_5 $ are given in \eqref{eq:beta}\textendash\eqref{eq:B5} shown at the bottom of this page, 
and let $ R = L+C $ be the overall number of paths or, equivalently, the tensor rank with path number $ C=PQ $ of the cascaded UE-RIS-BS channel. 
The process of converting the Kronecker product of vectors into a tensor is illustrated in Fig.~\ref{fig:tensor}.  
Let
$ \mathbf{B}_n = [\sqrt{x}\mathbf{1}_{1\times L} \otimes \mathbf{1}_{G_{n-1}\times1}, \mathbf{T}_n^H\mathbf{A}_{n} ], n=2,3 $ and 
$ \mathbf{B}_n = \mathbf{T}_n^H\mathbf{A}_{n}, n=4,5 $. 
Note that $ \{\mathbf{A}_{n}\}_{n=1}^5 $ are all Vandermonde matrices, but generally $\mathbf{A}_{1}$ is the merely Vandermonde CP factor of $ \bm{\mathcal{Y}} $ regardless of $ \{\mathbf{T}_{n}\}_{n=2}^5 $. 
In the noisy case, the received signal tensor is $ \hat{\bm{\mathcal{Y}}} = \bm{\mathcal{Y}} + \bm{\mathcal{W}} $ with noise tensor $ \bm{\mathcal{W}} $.

\begin{figure*}[hb]
	\centering
	\normalsize
	\setcounter{MYtempeqncnt}{\value{equation}}
	\vspace*{4pt}
	\hrulefill
	\begin{align}
		\label{eq:beta}
		\bm{\beta} &= \big[ \beta_{\textrm{L}}^{(1)},\ldots,\beta_{\textrm{L}}^{(L)},
		\beta_{\textrm{R}}^{(1,1)},\ldots,\beta_{\textrm{R}}^{(P,Q)} \big]^T \in\mathbb{C}^{R\times1},
		\\
		\label{eq:A1}
		\mathbf{A}_1 &= \big[ 
		\mathbf{a}^{(K)}(\omega^{(1)}_{1,\textrm{L}}),\ldots,
		\mathbf{a}^{(K)}(\omega^{(L)}_{1,\textrm{L}}),
		\mathbf{a}^{(K)}(\omega^{(1)}_{1,\textrm{R}}),\ldots,
		\mathbf{a}^{(K)}(\omega^{(P,Q)}_{1,\textrm{R}})
		\big] \in\mathbb{C}^{K\times R},
		\\
		\label{eq:B2}
		\mathbf{B}_2 &= \big[ 
		\underbrace{ \sqrt{x}\mathbf{1}_{G_1\times1},\ldots,
			\sqrt{x}\mathbf{1}_{G_1\times1} }_{L} ,
		\underbrace{ \mathbf{T}_2^H \mathbf{a}^{(M_y)}(\omega_2^{(1,1)}),\ldots,
			\mathbf{T}_2^H \mathbf{a}^{(M_y)}(\omega_2^{(P,Q)}) }_{C}
		\big] \in\mathbb{C}^{G_1\times R}  ,
		\\
		\label{eq:B3}
		\mathbf{B}_3 &= \big[ 
		\sqrt{x}\mathbf{1}_{G_2\times1},\ldots,
		\sqrt{x}\mathbf{1}_{G_2\times1} ,
		\mathbf{T}_3^H \mathbf{a}^{(M_z)}(\omega_3^{(1,1)}),\ldots,
		\mathbf{T}_3^H \mathbf{a}^{(M_z)}(\omega_3^{(P,Q)})
		\big] \in\mathbb{C}^{G_2\times R},
		\\
		\label{eq:B4}
		\mathbf{B}_4 &= \mathbf{T}_4^H \big[ 
		\underbrace{ \mathbf{a}^{(\tilde{N}_y)}(\omega_{4,\textrm{L}}^{(1)}),\ldots,
			\mathbf{a}^{(\tilde{N}_y)}(\omega_{4,\textrm{L}}^{(L)}) }_{L},
		\underbrace{ \mathbf{a}^{(\tilde{N}_y)}(\omega_{4,\textrm{R}}^{(1)}),\ldots,
			\mathbf{a}^{(\tilde{N}_y)}(\omega_{4,\textrm{R}}^{(1)}) }_{P} ,\ldots,
		\underbrace{ \mathbf{a}^{(\tilde{N}_y)}(\omega_{4,\textrm{R}}^{(Q)}),\ldots,
			\mathbf{a}^{(\tilde{N}_y)}(\omega_{4,\textrm{R}}^{(Q)}) }_{P}
		\big] \in\mathbb{C}^{N_1 \times R},
		\\
		\label{eq:B5}
		\mathbf{B}_5 &= \mathbf{T}_5^H \big[ 
		\mathbf{a}^{(\tilde{N}_z)}(\omega_{5,\textrm{L}}^{(1)}),\ldots,
		\mathbf{a}^{(\tilde{N}_z)}(\omega_{5,\textrm{L}}^{(L)}),
		\mathbf{a}^{(\tilde{N}_z)}(\omega_{5,\textrm{R}}^{(1)}),\ldots,
		\mathbf{a}^{(\tilde{N}_z)}(\omega_{5,\textrm{R}}^{(1)}) ,\ldots,
		\mathbf{a}^{(\tilde{N}_z)}(\omega_{5,\textrm{R}}^{(Q)}),\ldots,
		\mathbf{a}^{(\tilde{N}_z)}(\omega_{5,\textrm{R}}^{(Q)})
		\big] \in\mathbb{C}^{N_2 \times R}.
	\end{align}
	\setcounter{equation}{\value{equation}}
\end{figure*}

We analyze the CPD uniqueness of tensor $ \bm{\mathcal{Y}} $ concerning the classical Kruskal's condition~\cite[Sec.~3.2]{Kolda2009}. 
Since $ \mathbf{A}_1 $ is generally a full-rank Vandermonde matrix with distinct generators 
$ \{\omega_{\textrm{1,L}}^{(\ell)}\} \cup \{\omega_{\textrm{1,R}}^{(p,q)}\} $, 
$ k_{\mathbf{A}_1} = \operatorname{rank}(\mathbf{A}_1)=\min(K,R) $. 
Nevertheless, all the other factors $ \{\mathbf{B}_n\}_{n=2}^5 $ have repeated columns in the multipath scenario (i.e., $ L,P,Q \ge2 $), thus $ k_{\mathbf{B}_n}=1,\,n=2,3,4,5 $.  
As such, the Kruskal's condition for CPD uniqueness cannot be satisfied. To overcome the rank-deficient issue, spatial smoothing-based VSCPD technique will be employed to relax the uniqueness condition and acquire factor matrices. 

\subsection{ESPRIT Preliminaries}\label{sec:ESPRIT}

We briefly discuss the estimation of signal parameters via rotational invariance technique (ESPRIT) type mechanisms. 
The reader is referred to \cite[Sec.~III-B]{Zheng2024} for a detailed summary. 
ESPRIT provides an algebraic solution of generator $\bm{\omega}$ of Vandermonde matrix $ \mathbf{A} $. 
There are two estimation models: the element-space model and the transformed-space model.
In the element-space model, $ \mathbf{A} $ inherently has the shift-invariance property 
$ \mathbf{J}^{\uparrow}\mathbf{A}\operatorname{diag}(\bm{\omega}) = \mathbf{J}^{\downarrow}\mathbf{A} $ 
with selection matrices $ \{\mathbf{J}^{\uparrow},\mathbf{J}^{\downarrow}\} $. 
Once we obtain $ \hat{\mathbf{A}} $ satisfying 
$ \mathbf{A} = \hat{\mathbf{A}} \mathbf{D} $ with non-singular matrix $ \mathbf{D} $, 
$ \bm{\omega} $ and $ \mathbf{D} $ can be recovered from the eigenvalue decomposition (EVD) of 
$ (\mathbf{J}^{\uparrow}\hat{\mathbf{A}})^{\dagger} \mathbf{J}^{\downarrow}\hat{\mathbf{A}} $. 
In the transformed-space (e.g., beamspace) model, the signal subspace of $ \mathbf{A} $ is converted via 
$ \mathbf{B} = \mathbf{T}^H \mathbf{A} $ with transformation matrix $ \mathbf{T} $. 
Also, we can only acquire $ \hat{\mathbf{B}} $ with $ \mathbf{B} = \hat{\mathbf{B}} \mathbf{D} $. 
The profound works in \cite{Wen2018,Wen2020} have shown that we have similar shift-invariance property 
$ \mathbf{Q} \mathbf{B} \operatorname{diag}(\bm{\omega}) = \mathbf{Q} \mathbf{F}^H \mathbf{B} $ via preset projection matrix $ \mathbf{Q} $ and Vandermonde matrix $ \mathbf{T} $ satisfying 
$ \mathbf{J}^{\uparrow}\mathbf{T} = \mathbf{J}^{\downarrow}\mathbf{T}\mathbf{F} $. 
Subsequently, we can acquire $\bm{\omega}$ and $ \mathbf{D} $ through the EVD of 
$ (\mathbf{Q} \hat{\mathbf{B}})^{\dagger} \mathbf{Q} \mathbf{F}^H \hat{\mathbf{B}} $.

\subsection{VSCPD with Spatial Smoothing}

 The signal tensor is first augmented via spatial smoothing, depending on the Vandermonde structure. 
Since factor $ \mathbf{A}_1 $ is the only Vandermonde factor, we employ spatial smoothing in mode~$ 1 $, resulting in an extended signal tensor 
$ \bm{\mathcal{Y}}_{\textrm{sps}}\in\mathbb{C}^{K_1\times G_1\times G_2\times N_1\times N_2\times K_2 } $ with $ K_1+K_2-1=K $. 
The elements of $  \bm{\mathcal{Y}}_{\textrm{sps}} $ are expressed as 
\begin{align}
	&[\bm{\mathcal{Y}}_{\textrm{sps}}]_{k_1,g_1,g_2,n_1,n_2,k_2} 
	= [\bm{\mathcal{Y}}]_{k_1+k_2-1,g_1,g_2,n_1,n_2} \nonumber \\ 
	&= \sum_{r=1}^{R}
	[\bm{\beta}]_{r} e^{-\jmath 2 \pi (k_1-1)\Delta_f 
	[\bm{\tau}]_r} e^{-\jmath 2 \pi   (k_2-1)\Delta_f  [\bm{\tau}]_r}
	\nonumber \\
	&\quad\quad\;\;\times [\mathbf{B}_{2}]_{g_1,r} [\mathbf{B}_{3}]_{g_2,r} 
	[\mathbf{B}_{4}]_{n_1,r} [\mathbf{B}_{5}]_{n_2,r}, 
	\label{eq:sps}
\end{align}
where 
$ \bm{\tau}\in\mathbb{R}^{R\times1} $ is the delay collection defined like \eqref{eq:beta}. 
As such, $ \bm{\mathcal{Y}}_{\textrm{sps}} $ can be expressed in CPD form as 
\begin{equation}\label{eq:Ysps}
	\bm{\mathcal{Y}}_{\textrm{sps}} = \llbracket \bm{\beta};
	\mathbf{B}_1, \mathbf{B}_2, \mathbf{B}_3, \mathbf{B}_4, \mathbf{B}_5, \mathbf{B}_6 \rrbracket,
\end{equation}
where $ \mathbf{B}_n = \mathbf{T}_n^H \mathbf{A}_1, n=1,6 $ with two selection matrices 
$ \mathbf{T}_1^H = [\mathbf{I}_{K_1}, \mathbf{0}_{K_1\times (K-K_1)}] \in\mathbb{R}^{K_1\times K} $ and 
$ \mathbf{T}_6^H = [\mathbf{I}_{K_2}, \mathbf{0}_{K_2\times (K-K_2)}] \in\mathbb{R}^{K_2\times K} $. 
In the following, we give the general VSCPD uniqueness condition for a fifth-order tensor with only one Vandermonde factor. 
This is an extended version of \cite[Theorem~III.3]{Sorensen2013} to a higher-order case. 
\begin{theorem}\label{theo:unique}
	Let 
	$ \mathbf{A}_1\in\mathbb{C}^{K\times R} $, 
	$ \mathbf{B}_2\in\mathbb{C}^{G_1\times R} $, $ \mathbf{B}_3\in\mathbb{C}^{G_2\times R} $, 
	$ \mathbf{B}_4\in\mathbb{C}^{N_1\times R} $, and $ \mathbf{B}_5\in\mathbb{C}^{N_2\times R} $ 
	be the factor matrices of tensor  
	$ \bm{\mathcal{Y}} \in\mathbb{C}^{K\times G_1\times G_2\times N_1\times N _2} $ with weight vector $\bm{\beta}$, 
	and $ \mathbf{A}_1 $ is a Vandermonde matrix formed by distinct generators  
	$ [\bm{\omega}_1]_{r}, r=1,\ldots,R $. 
	Define $ \mathbf{B}_n = \mathbf{T}_n^H \mathbf{A}_1, n=1,6 $, where 
	$ \mathbf{T}_1^H = [\mathbf{I}_{K_1}, \mathbf{0}_{K_1\times (K-K_1)}] $ and 
	$ \mathbf{T}_6^H = [\mathbf{I}_{K_2}, \mathbf{0}_{K_2\times (K-K_2)}] $ 
	with $ K_1+K_2-1=K $. 
	If 
	\begin{equation}\label{eq:unique}
		\left\{\begin{array}{@{}l@{}}  
			\operatorname{rank}( \underline{\mathbf{B}}_1 \odot \mathbf{B}_2 \odot \mathbf{B}_3 ) = R \\
			\operatorname{rank}( \mathbf{B}_4 \odot \mathbf{B}_5 \odot \mathbf{B}_6 ) = R, 
		\end{array}\right.
	\end{equation}
\end{theorem}
then the VSCPD of $ \bm{\mathcal{Y}} $ is said to be unique.
Generically, \eqref{eq:unique} becomes 
\begin{equation}\label{eq:unique2}
	\min\big( (K_1-1)G_1G_2, K_2N_1N_2 \big) \ge R.
\end{equation}

\begin{IEEEproof}
	Following \cite[Theorem~3]{Jiang2001}, obtaining \eqref{eq:unique2} from \eqref{eq:unique} is straightforward. Consequently, we only concentrate on the proof of \eqref{eq:unique}. 
	Consider the extended tensor 
	$ \bm{\mathcal{Y}}_{\textrm{sps}} = \llbracket \bm{\beta};
	\mathbf{B}_1, \mathbf{B}_2, \mathbf{B}_3, \mathbf{B}_4, \mathbf{B}_5, \mathbf{B}_6 \rrbracket 
	\in\mathbb{C}^{K_1\times G_1\times G_2\times N_1\times N_2\times K_2 } $. 
	Denote the mode-$3$ matrix representation of $ \bm{\mathcal{Y}}_{\textrm{sps}} $ by 
	$ \mathbf{Y}_{\textrm{sps}, [3]} \in\mathbb{C}^{K_1G_1G_2\times N_1N_2K_2} $, expressed as  \cite[Eq.~(4)]{Sorensen2013} 
	\begin{equation}\label{eq:Y3}
		\mathbf{Y}_{\textrm{sps}, [3]} = 
		( \mathbf{B}_1 \odot \mathbf{B}_2 \odot \mathbf{B}_3 )
		\operatorname{diag}(\bm{\beta})
		( \mathbf{B}_4 \odot \mathbf{B}_5 \odot \mathbf{B}_6 )^{T}. 
	\end{equation}
	Define two selection matrices 
	\begin{align}\label{eq:J1J2}
		\mathbf{J}^{\uparrow} &=  
		[ \mathbf{I}_{K_1-1}, \mathbf{0}_{(K_1-1)\times1} ] \otimes
		\mathbf{I}_{G_1} \otimes \mathbf{I}_{G_2},  
		 \nonumber \\		
		\mathbf{J}^{\downarrow} &=  
		[ \mathbf{0}_{(K_1-1)\times1}, \mathbf{I}_{K_1-1} ] \otimes
		\mathbf{I}_{G_1} \otimes \mathbf{I}_{G_2} , 
	\end{align}
	then we come to 
	\begin{align}
		\mathbf{J}^{\uparrow} \mathbf{B}_{123} 
		&= \underline{\mathbf{B}}_1 \odot \mathbf{B}_2 \odot \mathbf{B}_3,     \nonumber \\
		\mathbf{J}^{\downarrow} \mathbf{B}_{123} 
		&= (\underline{\mathbf{B}}_1 \odot \mathbf{B}_2 \odot \mathbf{B}_3 )     \operatorname{diag}(\bm{\omega}_1), 
	\end{align}
	with $ \mathbf{B}_{123} = \mathbf{B}_1 \odot \mathbf{B}_2 \odot \mathbf{B}_3 $, and thus 
	\begin{equation}\label{eq:shift_in_B321}
		\mathbf{J}^{\uparrow} \mathbf{B}_{123} \operatorname{diag}(\bm{\omega}_1) = \mathbf{J}^{\downarrow} \mathbf{B}_{123}. 
	\end{equation}
	Given the conditions in \eqref{eq:unique}, we have the full-rank property 
	$ \operatorname{rank}\big(\mathbf{B}_{123}\big) = R $, and then following \eqref{eq:Y3} we also have 
	$ \operatorname{rank}\big(\mathbf{Y}_{\textrm{sps}, [3]}\big) = R $. 
	Let $ \mathbf{Y}_{\textrm{sps}, [3]} = 
	\mathbf{U} \mathbf{\Sigma} \mathbf{V}^H $ 
	be the compact singular value decomposition (SVD). 
	Notice that $ \mathbf{B}_{123} $ and $ \mathbf{U} $ share the same signal subspace, hence  
	\begin{equation}
		\mathbf{B}_{123} = \mathbf{U} \mathbf{D}, \;
		\mathbf{B}_{456} = \mathbf{V}^* \mathbf{\Sigma} \mathbf{D}^{-T} \operatorname{diag}(\bm{\beta})^{-1},  
	\end{equation}
	with $ \mathbf{B}_{456} = \mathbf{B}_4 \odot \mathbf{B}_5 \odot \mathbf{B}_6 $ and 
	non-singular matrix $ \mathbf{D} $. 
	Referring to Section~\ref{sec:ESPRIT}, we can utilize the EVD-based element-space ESPRIT to extract $ \bm{\omega}_1 $ and $ \mathbf{D} $. 
	Herein, \eqref{eq:unique} ensures the existence of 
	$ (	\mathbf{J}^{\uparrow} \mathbf{B}_{123})^{\dagger} $. 
	The ESPRIT results have permutation and scaling issues. i.e., 
	$ \hat{\bm{\omega}}_1^T = \bm{\omega}_1^T \mathbf{\Pi} $ and 
	$ \hat{\mathbf{D}} = \mathbf{D} \operatorname{diag}(\bm{\rho}) \mathbf{\Pi} $ 
	with permutation matrix $ \mathbf{\Pi} $ and scaling vector $ \bm{\rho} $. 
	Further, using generators, we can construct $ \hat{\mathbf{B}}_1 = \mathbf{B}_1\mathbf{\Pi} $ and $ \hat{\mathbf{B}}_6 = \mathbf{B}_6\mathbf{\Pi} $. 
	 
	Denote the $ r $th column of $ \mathbf{B}_{n}(\hat{\mathbf{B}}_{n}) $ by $ \mathbf{b}_{n,r}(\hat{\mathbf{b}}_{n,r}) $. 
	Further let $ \mathbf{b}_{23,r} = \mathbf{b}_{2,r}\otimes \mathbf{b}_{3,r} $ and 
	$ \mathbf{b}_{45,r} = \mathbf{b}_{4,r}\otimes \mathbf{b}_{5,r} $. 
	We have  
	\begin{align}
		\mathbf{b}_{23,r} &= K_1^{-1}
		 (\hat{\mathbf{b}}_{1,r}^H \otimes \mathbf{I}_{G_1G_2}  )
		 ( \hat{\mathbf{b}}_{1,r} \otimes \mathbf{b}_{23,r} ) ,
		 \label{eq:b32r} \\
		\mathbf{b}_{45,r} &= K_2^{-1}
		(\mathbf{I}_{N_1N_2} \otimes\hat{\mathbf{b}}_{6,r}^H  )
		( \mathbf{b}_{45,r} \otimes \hat{\mathbf{b}}_{6,r} ),
		\nonumber \\
		\hat{\mathbf{b}}_{1,r} \otimes \mathbf{b}_{23,r}& = 
		[\bm{\rho}]_r^{-1}\mathbf{U} \hat{\mathbf{d}}_r,\;
		\mathbf{b}_{45,r} \otimes \hat{\mathbf{b}}_{6,r} = 
		[\bm{\beta}]_r^{-1} [\bm{\rho}]_r \mathbf{V}^* \mathbf{\Sigma} \hat{\tilde{\mathbf{d}}}_r,
		\nonumber 
	\end{align}
	where $ \hat{\mathbf{d}}_{r}  (\hat{\tilde{\mathbf{d}}}_r) $ is the $r$th column of $ \hat{\mathbf{D}} (\hat{\mathbf{D}}^{-T}) $. 
	With \eqref{eq:b32r} at hand, we can solve $ \{\hat{\mathbf{b}}_{2,r}, \hat{\mathbf{b}}_{3,r}\} $ and $ \{\hat{\mathbf{b}}_{4,r}, \hat{\mathbf{b}}_{5,r}\} $ via rank-$1$ approximation, i.e., 
	\begin{align}
		&\!\!\!\min_{\hat{\mathbf{b}}_{2,r},\hat{\mathbf{b}}_{3,r}} \!\!\!\big\| 
		K_1\hat{\mathbf{b}}_{3,r}\hat{\mathbf{b}}_{2,r}^H -\operatorname{unvec}_{\scriptscriptstyle G_2 \times G_1}((\hat{\mathbf{b}}_{1,r}^H
		\otimes \mathbf{I}_{G_1G_2} ) \mathbf{U} \hat{\mathbf{d}}_r) \big\|_F^2,& 
		\nonumber \\
		&\!\!\!\min_{\hat{\mathbf{b}}_{4,r},\hat{\mathbf{b}}_{5,r}} \!\!\!\big\| 
		K_2\hat{\mathbf{b}}_{5,r}\hat{\mathbf{b}}_{4,r}^H -\operatorname{unvec}_{\scriptscriptstyle N_2 \times N_1}((\mathbf{I}_{N_1N_2} \otimes \hat{\mathbf{b}}_{6,r}^H ) \mathbf{V}^* \mathbf{\Sigma} \hat{\tilde{\mathbf{d}}}_r) \big\|_F^2.& 
		\nonumber
	\end{align}
	SVD can solve the problems above. The results of $ \hat{\mathbf{b}}_{3(5),r} $ and $ \hat{\mathbf{b}}_{2(4),r} $ are the left and right singular vectors of the second term inside the norm, respectively. 
\end{IEEEproof} 

The implementation of VSCPD with spatial smoothing is outlined in Algorithm~\ref{alg:VSCPD}. 
Importantly, our approach utilizes only standard linear algebra and avoids initialization and iterations by leveraging the power of the Vandermonde structure.  
Like other CPD techniques, VSCPD results still have permutation and scaling problems. 

\begin{algorithm}[t]
	\caption{VSCPD with Spatial Smoothing}
	\renewcommand{\algorithmicrequire}{\textbf{Input:}}
	\renewcommand{\algorithmicensure}{\textbf{Output:}}
	\label{alg:VSCPD}
	\begin{algorithmic}[1]
		\REQUIRE 
		Tensor $ \hat{\bm{\mathcal{Y}}} $ in the noisy case and rank $ R $. 
		\renewcommand{\algorithmicrequire}{\textbf{Implementation:}}
		\REQUIRE
		\STATE
		Apply spatial smoothing using \eqref{eq:sps} to form 
		$ \hat{\bm{\mathcal{Y}}}_{\textrm{sps}} $. 
		\STATE 
		Construct mode-$3$ matrix representation $ \hat{\mathbf{Y}}_{\textrm{sps}, [3]} $. 
		\STATE
		Calculate compact CPD 
		$ \hat{\mathbf{Y}}_{\textrm{sps}, [3]} = 
		\mathbf{U} \mathbf{\Sigma} \mathbf{V}^H $. 
		\STATE 
		Compute selection matrices $ \mathbf{J}^{\uparrow} $ and $ \mathbf{J}^{\downarrow} $ using \eqref{eq:J1J2}. 
		\STATE 
		Leverage $ \{ \mathbf{U},  \mathbf{J}^{\uparrow},  \mathbf{J}^{\downarrow} \} $ to acquire $ \{\hat{\bm{\omega}},  \hat{\mathbf{D}}\} $ using ESPRIT.  
		
		\FOR{$r=1$ to $R$}
		\STATE	
		Utilize generator $ [\hat{\bm{\omega}}_{1}]_r $ to construct $ \hat{\mathbf{b}}_{n,r},\, n=1,6 $. 
		\STATE
		Calculate SVD of 
		$ \operatorname{unvec}_{\scriptscriptstyle G_2\times G_1}
		((\hat{\mathbf{b}}_{1,r}^H \otimes \mathbf{I}_{G_1G_2} ) \mathbf{U} \hat{\mathbf{d}}_r) $ to obtain $ \hat{\mathbf{b}}_{n,r},\,n=2,3 $. 
		\STATE
		Compute SVD of 
		$ \operatorname{unvec}_{\scriptscriptstyle N_2\times N_1}
		( (\mathbf{I}_{N_1N_2} \otimes \hat{\mathbf{b}}_{6,r}^H )   
		\mathbf{V}^* \mathbf{\Sigma} \hat{\tilde{\mathbf{d}}}_r) $ to get  
		$ \hat{\mathbf{b}}_{n,r},\,n=4,5 $.
		\ENDFOR
		
		\ENSURE 
		Factors $ \{\hat{\mathbf{B}}_n\}_{n=1}^6 $ 
		and generator $ \hat{\bm{\omega}}_1 $. 
	\end{algorithmic}
\end{algorithm}

With Theorem~\ref{theo:unique}, we analyze the VSCPD uniqueness condition of our case with factor matrices shown in \eqref{eq:A1}\textendash\eqref{eq:B5}. 
We have 
$ k_{\underline{\mathbf{B}}_1} = \operatorname{rank}(\underline{\mathbf{B}}_1) = \min(K_1-1,R) $ with distinct generators in general. 
Recall that  
$ k_{\mathbf{B}_2} = k_{\mathbf{B}_3} =1 $ due to redundant columns in multipath scenario. 
For matrices $ \{\mathbf{X},\mathbf{Y}\} $ both having $R$ columns, we find  \cite{Jiang2001}
\begin{equation}
	\operatorname{rank} (\mathbf{X}\odot\mathbf{Y}) \ge k_{ \mathbf{X}\odot\mathbf{Y} } \ge 
	\min (k_{\mathbf{X}}+k_{\mathbf{Y}}-1,R). 
	\label{eq:rank_kr}
\end{equation}
Based on this property, 
$ \operatorname{rank}(\underline{\mathbf{B}}_1 \odot \mathbf{B}_2 \odot \mathbf{B}_3) = \min(K_1-1,R) $. 
To fulfill the first condition in \eqref{eq:unique}, it is necessary that $ K_1-1\ge R $. 
Similarly, to satisfy the second condition in \eqref{eq:unique}, it is necessary that  $ K_2\ge R $. 
In summary, the VSCPD is unique if 
\begin{equation}\label{eq:uniq_VSCPD}
	\min (K_1-1,K_2) \ge R,
\end{equation}
depending solely on spatial-smoothing parameters $\{K_1,K_2\}$ with given $R$. 
Importantly, \eqref{eq:uniq_VSCPD} provides a far more relaxed CPD uniqueness condition than the classical Kruskal's one.
With $ K_1+K_2-1=K $, the training subcarrier number should be set as $ K\ge2R $ in the frequency domain. 

The following subsections develop a triple-stage CE algorithm with factor estimates as the input. 
Stage~I performs multipath identification and algebraic coarse parameter estimation. Subsequently, Stages~II~and~III serve as optional double refinements at the cost of complexity growth.


\subsection{CE Stage I for Path Identification and Coarse Estimation}\label{sec:CEI}
Firstly, multipath identification is performed for different physical groups. Define $ \mathcal{R}=\{1,\ldots,R\} $, $ \mathcal{L}=\{1,\ldots,L\} $, $ \mathcal{P}=\{1,\ldots,P\} $, and $ \mathcal{Q}=\{1,\ldots,Q\} $.

Recall that $ \{\psi_2^{(p,q)},  \psi_3^{(p,q)}\} $ are only related to the cascaded UE-RIS-BS channel; thus we have to distinguish cascaded UE-RIS-BS paths from direct UE-BS paths. 
Thanks to our unique pilot design, \eqref{eq:B2} shows that the columns corresponding to direst UE-BS paths in $ \mathbf{B}_2 $ have the minimal $ L $ variances. 
The same is true for $ \mathbf{B}_3 $ in \eqref{eq:B3}. 
Therefore, we can deduce from $ \{\hat{\mathbf{b}}_{2(3),r}\}_{r=1}^R $ an index set $ \hat{\mathcal{L}}_{2(3)} $ for direct paths, associated with $ L $ lowest variance.  
This strategy is referred to as the {\it variance principle}, and here $ \hat{\mathcal{L}}_n \subset \mathcal{R}  $ with cardinality $ \operatorname{card}(\hat{\mathcal{L}}_n) = L $ for each $ n\in\{2,3\} $. 
At this moment, our pilot setting provides a chance to judge the algorithm status regarding tensor decomposition. 
In the successful decomposition case, we should have $ \hat{\mathcal{L}}_2 = \hat{\mathcal{L}}_3 $; otherwise, if $ \hat{\mathcal{L}}_2 \neq \hat{\mathcal{L}}_3 $ holds, algorithm fails and thus procedure aborts here.
To proceed, we will continue discussing the CE process for successful cases. 
Let $ \hat{\mathcal{L}} = \hat{\mathcal{L}}_2 $, and obtain the supplement 
$ \hat{\mathcal{C}} =\mathcal{R}\backslash \hat{\mathcal{L}} $ as the index set for cascaded paths 
with $ \operatorname{card}(\hat{\mathcal{C}}) = C$.

Further remind that $ \{\bm{\theta}_{\textrm{R}}^{(q)}\} $ solely come from the RIS-BS channel. 
Owing to the cascaded channel characteristic, each $ \omega_{4(5),\textrm{R}}^{(q)} $ originated from $ \bm{\theta}_{\textrm{R}}^{(q)} $, generates $ P $ duplicate columns in $ \mathbf{B}_{4(5)} $ (see \eqref{eq:B4}\textendash\eqref{eq:B5}). 
We hence aim to identify the cascaded paths formed by each RIS-BS sub-path using $ \{\hat{\mathbf{b}}_{n,r}\}_{r\in\hat{\mathcal{C}}},\,n\in\{4,5\} $. 
We define the correlation-based similarity criterion, i.e., 
\begin{equation}
	\operatorname{sim}( \hat{\mathbf{b}}_{n,r}, \hat{\mathbf{b}}_{n,r'} ) = 
	\frac{ \big| \hat{\mathbf{b}}_{n,r}^H \hat{\mathbf{b}}_{n,r'} \big| }
	{\|\hat{\mathbf{b}}_{n,r}\| \|\hat{\mathbf{b}}_{n,r'}\|} ,
	\; r\neq r'\in\hat{\mathcal{C}}. 
\end{equation}
With such a tool, we can group $ P $ paths that are similar enough to each other for $Q$ times, called {\it similarity principle}. 
That is, we obtain from $ \{\hat{\mathbf{b}}_{4(5),r}\}_{r\in\hat{\mathcal{C}}} $ an index set $ \hat{\mathcal{P}}^{(\hat{q})}_{4(5)} $ for each RIS-BS sub-path, $ \hat{q}\in\mathcal{Q} $. 
Herein, $ \operatorname{card}(\hat{\mathcal{P}}^{(\hat{q})}_{n})=P $ and $ \hat{\mathcal{P}}^{(1)}_{n}\cup\cdots\cup\hat{\mathcal{P}}^{(Q)}_{n} = \hat{\mathcal{C}} $ for every $ n\in\{4,5\} $. 
Again, we can determine the algorithm status here thanks to channel nature. 
If $ \hat{\mathcal{P}}^{(\hat{q})}_{4} = \hat{\mathcal{P}}^{(\hat{q})}_{5},\, \forall\hat{q}\in\mathcal{Q}$, tensor decomposition succeeds; otherwise the procedure breaks at this point. 
We continue presenting the subsequence in successful cases and let $ \hat{\mathcal{P}}^{(\hat{q})} =\hat{\mathcal{P}}^{(\hat{q})}_{4} $.

Using $ \{\hat{\mathbf{b}}_{n,r}\} $, we utilize the algebraic transformed-space ESRPIT introduced in Section~\ref{sec:ESPRIT} to recover generators in modes~$2\textendash5$. 
Nevertheless, instead of jointly calculating all generators like Section~\ref{sec:ESPRIT} in each mode, the column-wise scheme \cite{Wen2018} is leveraged, helping with the parameter association along different modes. 
That is,
\begin{equation}\label{eq:ESPRIT}
	\hat{\omega}_{n,r} = 
	(\mathbf{Q}_n \hat{\mathbf{b}}_{n,r})^{\dagger} \mathbf{Q}_n \mathbf{F}_n^H \hat{\mathbf{b}}_{n,r}, \;
	\left\{\begin{array} {@{}cc@{}} 
		r\in\hat{\mathcal{C}}, &\text{if}\; n\in\{2,3\}, \\
		r\in\mathcal{R}, &\text{if}\; n\in\{4,5\}.  
	\end{array}\right.
\end{equation}

The channel-parameter estimators using generator estimates are presented as follows. 
The VSCPD outputs $ \hat{\bm{\omega}}_1 $, and let $ \hat{\omega}_{1,r} = [\hat{\bm{\omega}}_1]_r,r\in\mathcal{R} $. 
The delay estimates are expressed as 
\begin{equation}\label{eq:taur}
	\hat{\tau}_r = - \frac{ \hat{\omega}_{1,r} }{2\pi\Delta_f}, \; r\in\mathcal{R}. 
\end{equation}
Then we conduct the element-wise mapping such that 
$ \hat{\tau}_r \mapsto \hat{\tau}^{ (\hat{\ell}) }_{\textrm{L}}, r \in\hat{\mathcal{L}},\hat{\ell}\in\mathcal{L} $ and 
$ \hat{\tau}_r \mapsto \hat{\tau}^{ (\hat{p},\hat{q}) }_{\textrm{R}},  r \in \hat{\mathcal{C}}, \hat{p}\in\mathcal{P} ,\hat{q}\in\mathcal{Q} $. 
In addition, we recover RIS's angle-related parameters from  
\begin{equation}\label{eq:psi_hat}
	\hat{\psi}_{n,r} = 
	\frac{\lambda}{2\pi d_{\textrm{R}}} \hat{\omega}_{n,r},
	\; n=2,3, r\in\hat{\mathcal{C}}.
\end{equation}
The mapping lies in  
$ \hat{\psi}_{1(2),r} \mapsto \hat{\psi}^{ (\hat{p},\hat{q}) }_{1(2)},  r \in \hat{\mathcal{C}},\hat{p}\in\mathcal{P} ,\hat{q}\in\mathcal{Q} $. 
Moreover, we acquire AOAs via 
\begin{equation}\label{eq:angle_est}
	\hat{\theta}_{\textrm{el},r} = 
	\arcsin \Big(  \frac{\lambda \hat{\omega}_{4,r} }{2\pi d_{\textrm{B}}} \Big), \;
	\hat{\theta}_{\textrm{az},r} = 
	\arcsin \Big(  \frac{\lambda \hat{\omega}_{3,r} }
	{ 2\pi d_{\textrm{B}}\cos(\hat{\theta}_{\textrm{el},r}) } \Big). 
\end{equation}
We then map such that 
$ \hat{\theta}_{\textrm{az(el)},r} \mapsto \hat{\theta}^{ (\hat{\ell}) }_{\textrm{L,az(el)}}  
, r \in \hat{\mathcal{L}}, \hat{\ell}\in\mathcal{L}  $ and 
$ \frac{1}{P} \sum_{ r\in\hat{\mathcal{P}}^{(\hat{q})} }\hat{\theta}_{\textrm{az(el)},r} \mapsto \hat{\theta}^{ (\hat{q}) }_{\textrm{R,az(el)}}, \hat{q}\in\mathcal{Q} $. 
Finally, we utilize $ \{\hat{\omega}_{n,r} \} $ to construct new factor estimates $ \{\hat{\mathbf{B}}_n\}_{n=1}^6 $. 
The path gains can be estimated via \cite[Eq.~(2.2)]{Kolda2009}
\begin{align}\label{eq:beta_est}
	\hat{\bm{\beta}} &=
	\mathbf{E}\,
	( \hat{\mathbf{B}}_6 \odot \hat{\mathbf{B}}_5 \odot \cdots \odot \hat{\mathbf{B}}_2 \odot \hat{\mathbf{B}}_1 )^H \operatorname{vec}(\bm{\mathcal{Y}}_{\textrm{sps}} ) ,
	\nonumber \\
	\mathbf{E} &= \big(
	(\hat{\mathbf{B}}_1^H \hat{\mathbf{B}}_1) \circledast
	(\hat{\mathbf{B}}_2^H \hat{\mathbf{B}}_2) \circledast\cdots\circledast
	 (\hat{\mathbf{B}}_6^H \hat{\mathbf{B}}_6) 
	 \big)^{\dagger} . 
\end{align} 
Let $ [\hat{\bm{\beta}}]_r =\hat{\beta}_r,r\in\mathcal{R} $. 
We map that 
$ \hat{\beta}_r \mapsto \hat{\beta}^{ (\hat{\ell}) }_{\textrm{L}}, r \in\hat{\mathcal{L}},\hat{\ell}\in\mathcal{L} $ and 
$ \hat{\beta}_r \mapsto \hat{\beta}^{ (\hat{p},\hat{q}) }_{\textrm{R}},  r \in \hat{\mathcal{C}}, \hat{p}\in\mathcal{P} ,\hat{q}\in\mathcal{Q} $. 
Algorithm~\ref{alg:CE} concludes our CE algorithm in Stage~I.

\begin{algorithm}[t]
	\caption{Multipath Identification and Algebraic CE Algorithm (Stage~I)}
	\renewcommand{\algorithmicrequire}{\textbf{Input:}}
	\renewcommand{\algorithmicensure}{\textbf{Output:}}
	\label{alg:CE}
	\begin{algorithmic}[1]
		\REQUIRE 
		Factor columns $ \{\hat{\mathbf{b}}_{n,r}\} $, generator $ \hat{\bm{\omega}}_1 $, and path numbers $ \{L, P, Q \} $. 

		\renewcommand{\algorithmicrequire}{\textbf{Implementation:}}
		\REQUIRE
		
		\STATE 
		Determine $ \hat{\mathcal{L}}_n, n\in\{2,3\} $ based on {\it variance principle}. 
		\IF{ {\it $ \hat{\mathcal{L}}_2 \neq \hat{\mathcal{L}}_3 $ } }
		\STATE 
		Algorithm fails and \textbf{break}. 
		\ELSE 
		\STATE
		Let $ \hat{\mathcal{L}} = \hat{\mathcal{L}}_2 $ and determine its supplement $ \hat{\mathcal{C}} = \mathcal{R}\backslash \hat{\mathcal{L}} $. 
		\STATE
		Find $ \{ \hat{\mathcal{P}}^{(\hat{q})}_n \}_{\hat{q}=1}^Q,  n\in\{4,5\} $ using {\it similarity principle}. 
		\IF{ {\it $ \hat{\mathcal{P}}^{(\hat{q})}_4 \neq \hat{\mathcal{P}}^{(\hat{q})}_5, \exists \hat{q}\in\mathcal{Q} $ } }
		\STATE
		Algorithm fails and \textbf{break}. 
		\ELSE
		\STATE
		Let $\hat{\mathcal{P}}^{(\hat{q})} = \hat{\mathcal{P}}^{(\hat{q})}_4 $. 
		\FOR{ $ r\in\hat{\mathcal{C}} $ if $ n\in\{2,3\} $ \textbf{or} $ r\in\mathcal{R} $ if $ n\in\{4,5\} $ }
		\STATE
		Calculate $ \hat{\omega}_{n,r} $ for each $ n $  using \eqref{eq:ESPRIT}.  
		\ENDFOR
		\STATE 
		Recover multipath parameters using \eqref{eq:taur}\textendash\eqref{eq:beta_est}.
		\ENDIF
		\ENDIF		
		\ENSURE Channel-parameter estimates  
		and sets 
		$ \{ \hat{\mathcal{L}},\hat{\mathcal{C}}, \hat{\mathcal{P}}^{(\hat{q})} \} $.   
	\end{algorithmic}
\end{algorithm}

We have treated the path numbers $ \{R,L,P,Q\} $ as known values. 
In the case without such prior information, we can also estimate these path numbers under our framework. 
First, $\hat{R}$ can be determined by the minimum description length (MDL) technique \cite{Wax1985}. 
Further, in $ \{\hat{\mathbf{b}}_{2(3),r}\}_{r=1}^{\hat{R}} $, the lowest $L $ variances should be much smaller than the lowest $ (L+1) $th one; based on such result, we can extend our {\it variance principle} to obtain $ \hat{L} $. 
Analogously, $ \{\hat{P},\hat{Q}\} $ can be also inferred using the extended {\it similarity principle}.

\begin{algorithm}[t]
	\caption{Search-Based CE Refinement Method (Stage~II)}
	\renewcommand{\algorithmicrequire}{\textbf{Input:}}
	\renewcommand{\algorithmicensure}{\textbf{Output:}}
	\label{alg:CE2}
	\begin{algorithmic}[1]
		\REQUIRE 
		Coarse parameter estimates, factors $ \{\hat{\mathbf{B}}_n\}_{n=2}^5 $, matrices $ \{\mathbf{T}_n\}_{n=2}^5 $, index sets  $ \{ \hat{\mathcal{L}},\hat{\mathcal{C}}, \hat{\mathcal{P}}^{(\hat{q})}  \} $, 
		and search parameters $ \{ \Delta_n^1, E_n, I_n, \zeta_n \}_{n=2}^5 $.
		\renewcommand{\algorithmicrequire}{\textbf{Implementation:}}
		\REQUIRE
		
		\FOR{ $n\in\{2,3,4,5\}$ }		
		\FOR{ $ r\in\hat{\mathcal{C}} $ if $ n\in\{2,3\} $ 
			\textbf{or} $ r\in\mathcal{R} $ if $ n\in\{4,5\} $ }
		\FOR{ $i=1$ to $I_n$ }
		\STATE 
		Construct search space $ \mathcal{D}_{n,r}^i $ using \eqref{eq:D_nir}. 
		\STATE
		Search in $ \mathcal{D}_{n,r}^i $ to find a $ \hat{\omega}_{n,r} $ maximizing  \eqref{eq:CBS}. 
		\STATE
		Update search resolution $ \Delta_n^i $ using \eqref{eq:Delta_ni}. 
		\ENDFOR
		\ENDFOR
		\ENDFOR
		
		\STATE
		Obtain multipath parameters using \eqref{eq:psi_hat}\textendash\eqref{eq:beta_est}.
		
		\ENSURE Refined Channel-parameter estimates.   
	\end{algorithmic}
\end{algorithm}

\subsection{CE Stage II for Estimate Refinement }

Although Stage~I provides ESPRIT-based algebraic solution with fast implementation, its output is not accurate enough \cite{Zheng2024}.  To remedy this, we apply the correlation-based search (CBS) to improve the accuracy. 
Specifically, the optimization problem for only one variable is given by 
\begin{equation}\label{eq:CBS}
	\max_{\hat{\omega}_{n,r}} 
	 \frac{ \big| \hat{\mathbf{b}}_{n,r}^H \mathbf{b}(\hat{\omega}_{n,r}) \big| } 
	{ \|\hat{\mathbf{b}}_{n,r}\|  \|\mathbf{b}(\hat{\omega}_{n,r})\| }, \;
	\left\{\begin{array} {@{}cc@{}} 
		r\in\hat{\mathcal{C}}, &\text{if}\; n\in\{2,3\}, \\
		r\in\mathcal{R}, &\text{if}\; n\in\{4,5\},  
	\end{array}\right.
\end{equation}
where $ \mathbf{b}(\hat{\omega}_{n,r}) = 
\mathbf{T}_n^H \mathbf{a}^{(X)}(\hat{\omega}_{n,r}),X\in\{M_y,M_z,\tilde{N}_y,\tilde{N}_z\} $ corresponding to each $n\in\{2,3,4,5\}$. 
Per \cite[Appendix~A]{Zhou2017}, the CBS scheme is equivalent to the maximum likelihood (ML) estimator provided that the estimation error of $ \hat{\mathbf{b}}_{n,r} $ follows the circularly symmetric Gaussian distribution. 
Nonetheless, instead of searching exhaustively to solve \eqref{eq:CBS} like \cite{Zhou2017,Zheng2022,He2024},  the following iterative search strategy is developed here. 
Let $ \Delta^{i}_{n},n\in\{2,3,4,5\} $ be the $i$th search resolution, $ i=1,\ldots,I_n $. 
Further, let $ \mathcal{D}^{i}_{n,r} $ denote the $i$th search space centered at the estimate obtained from the last iteration (input the estimate obtained in Stage I as the initial value). 
That is, $ \mathcal{D}^{i}_{n,r} $ is constructed from  
\begin{equation}\label{eq:D_nir}
	\mathcal{D}^{i}_{n,r} = \{
	\ldots, \hat{\omega}_{n,r}^{i-1}-\Delta^{i}_{n}, \hat{\omega}_{n,r}^{i-1}, \hat{\omega}_{n,r}^{i-1}+\Delta^{i}_{n}, \ldots  \}.
\end{equation} 
We fix the cardinality $ \operatorname{card} (\mathcal{D}^{i}_{n,r}) = E_n $. 
During each iteration, search in $  \mathcal{D}^{i}_{n,r} $ and output the estimate maximizing \eqref{eq:CBS}. 
Then resolution $ \Delta^{i}_{n} $ is shrunk in the following manner
\begin{equation}\label{eq:Delta_ni}
	\Delta^{i+1}_{n} = \zeta_{n} \Delta^{i}_{n},\; i =1,\ldots,I_n-1,
\end{equation}
with coefficient $ 0<\zeta_ {n}<1 $. 
Repeating the iteration $ I_n $ times yields CBS-refined estimates $ \{\hat{\omega}_{n,r}\} $.

Subsequently,  the updated $ \{\hat{\omega}_{n,r}\} $ can be used to reconstruct channel parameters (except delays) with better accuracy via \eqref{eq:psi_hat}\textendash\eqref{eq:beta_est}. 
Algorithm~\ref{alg:CE2} summarizes our CE Stage~II.

\subsection{CE Stage III for Further Refinement} 

The factor matrix estimates obtained in Stage~II deliver a reliable CPD result. 
Such result can be then used as the initialization for the iterative ALS algorithm to produce more precise CP factors.
In particular, factor estimators can be constructed by solving the following problem 
\begin{equation}\label{eq:ALS}
	 \min_{\{\hat{\mathbf{B}}_n\}_{n=1}^6} \big\| 
	\bm{\mathcal{Y}}_{\textrm{sps}} - 
	\llbracket 
	\mathbf{1}_R;
	\hat{\mathbf{B}}_1, \hat{\mathbf{B}}_2, \hat{\mathbf{B}}_3, \hat{\mathbf{B}}_4, \hat{\mathbf{B}}_5, \hat{\mathbf{B}}_6 \rrbracket \big\|_F^2, 
\end{equation}
where  weight $ \hat{\bm{\beta}} $ is absorbed into any factor, e.g., 
$ \hat{\mathbf{B}}_6 \leftarrow \hat{\mathbf{B}}_6\operatorname{diag}(\hat{\bm{\beta}}) $. 
The ALS scheme alternately updates factor $ \hat{\mathbf{B}}_n $ by fixing the other factors and computing the least squares (LS) estimate as 
\begin{align}\label{eq:ALS_Bn}
	&\!\!\!\hat{\mathbf{B}}_n = 
	\mathbf{Y}_{\textrm{sps},(n)} ( \hat{\mathbf{B}}_6 \odot\cdots \hat{\mathbf{B}}_{n+1} \odot \hat{\mathbf{B}}_{n-1} \cdots \odot \hat{\mathbf{B}}_{1} )^* \mathbf{F}^*,
	\\
	&\!\!\!\mathbf{F} = 
	( \hat{\mathbf{B}}_{1}^H\hat{\mathbf{B}}_{1} \odot\cdots 
	\hat{\mathbf{B}}_{n-1}^H\hat{\mathbf{B}}_{n-1} \odot 
	\hat{\mathbf{B}}_{n+1}^H\hat{\mathbf{B}}_{n+1} \odot \cdots
	\hat{\mathbf{B}}_6^H\hat{\mathbf{B}}_6 )^{\dagger}, \nonumber
\end{align}
where $ \mathbf{Y}_{\textrm{sps},(n)} $ is the mode-$n$ matricization of $ \bm{\mathcal{Y}}_{\textrm{sps}} $ formed by arranging the mode-$n$ fibers to be the columns of the resulting matrix \cite{Kolda2009}.

The factor estimates $ {\{\hat{\mathbf{B}}_n\}_{n=1}^6} $
  obtained in Stage~II can be applied to initialize ALS. Once ALS converges under the preset threshold, new CP factors are acquired. Algorithms~\ref{alg:CE}~and~\ref{alg:CE2}  can then be repeated to refine the channel-parameter estimates further. To update the delay estimates, column-wise element-space ESPRIT needs to be performed using the new $ \hat{\mathbf{B}}_1 $ and/or $ \hat{\mathbf{B}}_6 $, similar to the implementation of transformed-space ESPRIT in \eqref{eq:ESPRIT}. If both modes~1 and 6 are selected, the corresponding delay estimates are averaged and the mean value is output, as there is no clear indication of which is more accurate. It is observed that element-space ESPRIT alone suffices to ensure the accuracy of delay estimates, making the subsequent CBS refinement unnecessary. CE Stage~III is outlined in Algorithm~\ref{alg:CE3}.

\begin{algorithm}[t]
	\caption{Iterative CE Refinement Method (Stage~III)}
	\renewcommand{\algorithmicrequire}{\textbf{Input:}}
	\renewcommand{\algorithmicensure}{\textbf{Output:}}
	\label{alg:CE3}
	\begin{algorithmic}[1]
		\REQUIRE 
		Coarse factor estimates, noisy tensor $ \hat{\bm{\mathcal{Y}}}_{\textrm{sps}} $, 
		maximum iteration number $ I_{\textrm{max}} $, and convergence threshold $ \epsilon $.  
		\renewcommand{\algorithmicrequire}{\textbf{Implementation:}}
		\REQUIRE
		
		\STATE
		Let $ \hat{\bm{\mathcal{Y}}}_{\textrm{sps}}^0 = \hat{\bm{\mathcal{Y}}}_{\textrm{sps}} $ for initialization. 
		\FOR{ $ i=1 $ to $ I_{\textrm{max}} $ }		
			\FOR{ $ n=1 $ to $ 6 $ }
				\STATE 
				Construct mode-$n$ matricization $ \hat{\mathbf{Y}}_{\textrm{sps},(n)} $ of $ \hat{\bm{\mathcal{Y}}}_{\textrm{sps}} $. 
				\STATE
				Update factor estimate $ \hat{\mathbf{B}}_n $ using \eqref{eq:ALS_Bn}.		
			\ENDFOR
			\STATE 
			Obtain tensor estimate
			$ \hat{\bm{\mathcal{Y}}}_{\textrm{sps}}^{i} = \llbracket \mathbf{1}_R; 
			\hat{\mathbf{B}}_1, \hat{\mathbf{B}}_2, \ldots, \hat{\mathbf{B}}_6 \rrbracket   $. 
			\IF{ 
				$   \| \hat{\bm{\mathcal{Y}}}_{\textrm{sps}}^i -
					 \hat{\bm{\mathcal{Y}}}_{\textrm{sps}}^{i-1} \|_F / 
					  \| \hat{\bm{\mathcal{Y}}}_{\textrm{sps}}^i  \|_F < \epsilon   $  }
				\STATE
				\textbf{break.}
			\ENDIF
		\ENDFOR
		
		\STATE
		Perform Algorithms~\ref{alg:CE} and \ref{alg:CE2}.
		
		\ENSURE Refined Channel-parameter estimates.   
	\end{algorithmic}
\end{algorithm}

\section{Performance Analysis}\label{sec:PA}

\subsection{CRLB Analysis}\label{sec:CRLB}

In this subsection, the CRLB is determined to indicate the achievable optimal estimation precision. Note that CRLB completely depends on the observation model, irrelevant to the specific estimator \cite{Sengupta1993}, e.g., either matrix-based or tensor-based solution. The CRLB is calculated from the inverse of the Fisher information matrix (FIM). In this paper, FIM is constructed from the real-valued observation model by transforming the original complex-valued observation model into a real-valued one, i.e., partitioning the real and imaginary parts of the complex numbers. Before discussing the specific case, the FIM for the general real Gaussian scenario is first considered.

\begin{theorem}\label{theo:FIM}
	Assume that 
	$ \mathbf{x} \sim \mathcal{N} \big( \bm{\mu}(\bm{\phi}), \mathbf{C}(\bm{\phi}) \big) $, 
	and hence both the mean $  \bm{\mu}(\bm{\phi}) $ and covariance $ \mathbf{C}(\bm{\phi}) $ may depend on parameter $ \bm{\phi} $. 
	Let $ [\bm{\phi}]_i = \phi_i $, 
	then the FIM  is given by 
 	\begin{align}\label{eq:FIM}
 		{[\mathbf{J}(\bm{\phi})]_{i, j} }  =
 		&\left[ \frac{\partial \bm{\mu}(\bm{\phi})} {\partial \phi_{i}} \right]^{T} 
 		 \mathbf{C}^{-1}(\bm{\phi})
 		\left[ \frac{\partial \bm{\mu}(\bm{\phi})}{\partial \phi_{j}} \right] 
 		\nonumber \\
 		& +\frac{1}{2} \operatorname{tr}\left[ 
 		\mathbf{C}^{-1}(\bm{\phi}) \frac{\partial \mathbf{C}(\bm{\phi})}{\partial \phi_{i}} 
 		\mathbf{C}^{-1}(\bm{\phi}) \frac{\partial \mathbf{C}(\bm{\phi})}{\partial \phi_{j}} \right].
 	\end{align}
\end{theorem}

\begin{IEEEproof}
	Details can be found in \cite[Appendix~3C]{Sengupta1993}. 
\end{IEEEproof}


For the case at hand, the real observation model can be expressed as
\begin{equation}\label{eq:observation}
	\hat{ \tilde{\mathbf{y}} }^{(k,g)} = \tilde{\mathbf{y}}^{(k,g)} (\bm{\phi}) + \tilde{\mathbf{w}}^{(k,g)}, 
	k=1,\ldots,K, g=1,\ldots,G,
\end{equation}
where 
$ \tilde{\mathbf{y}}^{(k,g)} = 
\big[ \mathfrak{R}(\mathbf{y}^{(k,g)})^T, \mathfrak{I}(\mathbf{y}^{(k,g)})^T  \big]^T 
\in \mathbb{C}^{ 2N_1N_2\times1  } $ with 
$ \mathbf{y}^{(k,g)} = \mathbf{y}_{\textrm{L}}^{(k,g)} + \mathbf{y}_{\textrm{R}}^{(k,g)} $ 
being noise-free ensemble signal in \eqref{eq:ykg2},  
$ \tilde{\mathbf{w}}^{(k,g)} = 
\big[ \mathfrak{R}(\mathbf{w}^{(k,g)})^T, \mathfrak{I}(\mathbf{w}^{(k,g)})^T  \big]^T  $, 
and parameter vector $ \bm{\phi} $ will be defined later.  

Following \cite[Appendix~A]{Zheng2024}, the covariance matrix of $ \tilde{\mathbf{w}}^{(k,g)} $ is 
$ \mathbf{C}^{(k,g)} = \mathbf{C}^{(k,g)}_{\textrm{B}} + \mathbf{C}^{(k,g)}_{\textrm{R}} $ with 
\begin{equation}
	 \mathbf{C}^{(k,g)}_{\textrm{X}} = 
	 \frac{ \sigma^2_{\textrm{X}} }{2} \!\!
	\begin{array}{l}
		\setlength{\arraycolsep}{0.5pt}
		\left[ \begin{array}{@{}cc@{}}
			\mathfrak{R}( \mathbf{G}_{\textrm{X}}\mathbf{G}_{\textrm{X}}^H ) & 
			\mathfrak{I}( \mathbf{G}_{\textrm{X}}\mathbf{G}_{\textrm{X}}^H )^T  \\ 
			\mathfrak{I}( \mathbf{G}_{\textrm{X}}\mathbf{G}_{\textrm{X}}^H ) & 
			\mathfrak{R}( \mathbf{G}_{\textrm{X}}\mathbf{G}_{\textrm{X}}^H )
		\end{array} \right], 
	\end{array}
\end{equation}
where $ \textrm{X}\in\{\textrm{B},\textrm{R}\} $, 
$ \mathbf{G}_{\textrm{B}} = \mathbf{R}^H $, and
$ \mathbf{G}_{\textrm{R}} = \mathbf{R}^H \mathbf{H}_{\textrm{R},2} ^{(k)} \mathbf{\Gamma}^{(g)} $. 
We see that in our active-RIS scenario here, the covariance matrix $ \mathbf{C}^{(k,g)} $ is a function of $ \mathbf{H}_{\textrm{R},2} ^{(k)} $ and thus relevant to channel parameters. 
However, in the passive-RIS case without thermal noise at RIS (e.g., see \cite{Wei2021,Xu2022,Huang2022}), the noise covariance (i.e., $ \mathbf{C}^{(k,g)}=\mathbf{C}^{(k,g)}_{\textrm{B}} $) does not depend on channel parameters, hence the second term in \eqref{eq:FIM} is zero, which is a special case of ours. 
Note that the active-RIS work in \cite{Zheng2024} has neglected the second term in \eqref{eq:FIM}.

Using \eqref{eq:HR2}, $ \mathbf{G}_{\textrm{R}}\mathbf{G}_{\textrm{R}}^H $ can be further expressed as 
\begin{align}
	\mathbf{G}_{\textrm{R}}&\mathbf{G}_{\textrm{R}}^H = 
	\eta^2 M_y M_z
	\sum_{q} \big|\beta_{\textrm{R},2}^{(q)}\big|^2 
	\mathbf{R}^H 
	\mathbf{A}_{\textrm{B}} \big( \bm{\theta}_{\textrm{R}}^{(q)} \big)
	\mathbf{R} 
	\nonumber \\  
	&+ \eta^2 \sum_{q\neq q'} 
	\beta_{\textrm{R},2}^{(q)} (\beta_{\textrm{R},2}^{(q')})^*
	e^{ \jmath 2 \pi f^{(k)} ( \tau_{\textrm{R},2}^{(q')}-\tau_{\textrm{R},2}^{(q)} ) } 
	\mathbf{R}^H 
	\bar{\mathbf{A}}_{\textrm{B}} 
	\mathbf{R}, 
	\label{eq:GG^H}
\end{align}
where we have leveraged  
$ \mathbf{\Gamma}^{(g)} (\mathbf{\Gamma}^{(g)})^H = \eta^2\mathbf{I}_{M_yM_z} $ as well as  
$ 	\mathbf{a}_{\textrm{R}}^T \big( \bm{\varphi}_{\textrm{D}}^{(q)} \big)
	\mathbf{a}_{\textrm{R}}^* \big( \bm{\varphi}_{\textrm{D}}^{(q)} \big) = M_yM_z $, and 
\begin{gather}
	\mathbf{A}_{\textrm{B}} \big( \bm{\theta}_{\textrm{R}}^{(q)} \big) = 
	\mathbf{a}_{\textrm{B}} \big( \bm{\theta}_{\textrm{R}}^{(q)} \big)
	\mathbf{a}_{\textrm{B}}^H \big( \bm{\theta}_{\textrm{R}}^{(q)} \big) 
    \in \mathbb{C}^{\tilde{N}_y\tilde{N}_z\times\tilde{N}_y\tilde{N}_z},  
	\nonumber \\
	\bar{\mathbf{A}}_{\textrm{B}}
	=
	\big( \mathbf{a}_{\textrm{R}}^T \big( \bm{\varphi}_{\textrm{D}}^{(q)} \big)
	\mathbf{a}_{\textrm{R}}^* \big( \bm{\varphi}_{\textrm{D}}^{(q')} \big) \big)
	\mathbf{a}_{\textrm{B}} \big( \bm{\theta}_{\textrm{R}}^{(q)} \big)
	\mathbf{a}_{\textrm{B}}^H \big( \bm{\theta}_{\textrm{R}}^{(q')} \big). 
\end{gather}
Note in \eqref{eq:GG^H} that the first term is greatly larger than the second term, owing to aligned phases in the first term. Therefore, it is reasonable to neglect the second term in \eqref{eq:GG^H} hereafter. 

We determine the unknowns for CRLB calculation as follows. 
Firstly, the multipath parameters of interest (aiming to recover direct and cascaded channels) were listed at the end of Section~\ref{sec:SM}.
Define $ \bm{\tau}_{\textrm{L}}\in\mathbb{R}^{L\times1} $, $ \bm{\tau}_{\textrm{R}}\in\mathbb{R}^{C\times1} $, $ \bm{\psi}_{2(3)}\in\mathbb{R}^{C\times1} $, $ \bm{\theta}_{\textrm{L}}\in\mathbb{R}^{2L\times1} $, $ \bm{\theta}_{\textrm{R}}\in\mathbb{R}^{2Q\times1} $, $ \bm{\beta}_{\textrm{L}}\in\mathbb{R}^{L\times1} $, and $ \bm{\beta}_{\textrm{R}}\in\mathbb{R}^{C\times1} $ as the vector collection of $ \{\tau_{\textrm{L}}^{(\ell)}\} $, $ \{\tau_{\textrm{R}}^{(p,q)}\} $, $ \{\psi_{2(3)}^{(p,q)}\} $, $ \{\theta_{\textrm{L,az}}^{(\ell)},\theta_{\textrm{L,el}}^{(\ell)}\} $, $ \{\theta_{\textrm{R,az}}^{(q)},\theta_{\textrm{R,el}}^{(q)}\} $, $ \{\beta_{\textrm{L}}^{(\ell)}\} $, and $ \{\beta_{\textrm{R}}^{(p,q)}\} $, respectively. 
Let $ \bm{\beta}_{\textrm{L(R)},\mathfrak{R}} = \mathfrak{R}(\bm{\beta}_{\textrm{L(R)}}) $ and $ \bm{\beta}_{\textrm{L(R)},\mathfrak{I}} = \mathfrak{I}(\bm{\beta}_{\textrm{L(R)}}) $. 
Then we can form parameter vector of interest of dimension $ 5R+2Q $ as
\begin{equation}
	\bm{\phi}_{\textrm{int}} =
	[ \bm{\tau}_{\textrm{L}}^T, \bm{\tau}_{\textrm{R}}^T, \bm{\psi}_{2}^T, \bm{\psi}_{3}^T, \bm{\theta}_{\textrm{L}}^T, \bm{\theta}_{\textrm{R}}^T, \bm{\beta}_{\textrm{L},\mathfrak{R}}^T, \bm{\beta}_{\textrm{L},\mathfrak{I}}^T, \bm{\beta}_{\textrm{R},\mathfrak{R}}^T, \bm{\beta}_{\textrm{R},\mathfrak{I}}^T ]^T. \nonumber
\end{equation} 
Secondly, the nuisance parameters generated from the covariance term in \eqref{eq:GG^H} construct the vector 
$ \bm{\phi}_{\textrm{nui}} = \big[ 
\big|\beta_{\textrm{R},2}^{(1)}\big|^2, \ldots, \big|\beta_{\textrm{R},2}^{(Q)}\big|^2 \big]^T
\in \mathbb{R}^{Q\times1} $. 
Consequently, the overall parameter vector becomes
$ \bm{\phi} = [\bm{\phi}_{\textrm{int}}^T, \bm{\phi}_{\textrm{nui}}^T]^T \in \mathbb{R}^{(5R+3Q)\times1}  $. 

With \eqref{eq:observation}, we have  
$ \hat{ \tilde{\mathbf{y}} }^{(k,g)} \sim \mathcal{N} 
\big( \tilde{\mathbf{y}}^{(k,g)} (\bm{\phi}), \mathbf{C}^{(k,g)}(\bm{\phi}) \big) $. 
Based on Theorem~\ref{theo:FIM}, we can compute the FIM 
$ \mathbf{J}^{(k,g)}(\bm{\phi}) $ for each $ k\in\{1,\ldots,K\} $ and each $ g\in\{1,\ldots,G\} $. 
The partial derivatives in FIM calculation are concluded in Appendix~\ref{sec:Derivatives}.  
Thus, the ensemble FIM across all training subcarriers and time slots is expressed as 
\begin{equation}
	\mathbf{J}(\bm{\phi}) = \sum_{k=1}^{K} \sum_{g=1}^{G} \mathbf{J}^{(k,g)}(\bm{\phi})  
	\in\mathbb{R}^{\scriptscriptstyle(5R+3Q)\times(5R+3Q)}.
\end{equation}
Our interest is to determine the performance bound for  $ \bm{\phi}_{\textrm{int}} $. Therefore, invoke the inverse formula of a partitioned matrix \cite{Horn2012} for the $ (5R+2Q)\times (5R+2Q) $ upper left block, yielding the equivalent FIM for $ \bm{\phi}_{\textrm{int}} $ only as 
\begin{equation} 
	\breve{\mathbf{J}}(\bm{\phi}_{\textrm{int}}) = 
	\mathbf{J}_{\textrm{1}} - 
	\mathbf{J}_{\textrm{2}} \mathbf{J}_{\textrm{3}}^{-1} \mathbf{J}_{\textrm{2}}^T, \;
	\mathbf{J}(\bm{\phi}) =  \!\!
	\begin{array}{l}
		\setlength{\arraycolsep}{0.5pt}
		\left[ \begin{array}{cc} 
			\mathbf{J}_{\textrm{1}} & 
			\mathbf{J}_{\textrm{2}}  \\ 
			\mathbf{J}_{\textrm{2}}^T & 
			\mathbf{J}_{\textrm{3}}
		\end{array} \right]. 
	\end{array}
\end{equation}
The CRLBs for elements in $ \bm{\phi}_{\textrm{int}} $ hold 
\begin{equation}\label{}
	\operatorname{CRLB} \left( [\bm{\phi}_{\textrm{int}}]_s \right) = 
	\big[ \breve{\mathbf{J}}^{-1}(\bm{\phi}_{\textrm{int}}) \big]_{s,s}, 
	\ s=1,\ldots,5R+2Q. 
\end{equation}

\subsection{Complexity Analysis}

The complexity of the  proposed algorithms and the benchmarking methods is analyzed next. 
To simplify the analysis, it is set that 
$ K_1 = K_2 $, $ G_1=G_2 $, and $ N_1 = N_2 $; thus,  their subscripts $ 1 $ and $ 2 $ are omitted in this subsection. The CPDs in baseline methods are realized via ALS, and all ALS mechanisms require $ T $ iterations. 

Our main CE framework can be divided into four parts (Fig.~\ref{fig:flow_chart}): VSCPD and CE Stages~I\textendash III. 
Let us start with the complexity of VSCPD. 
The compact SVD of 
$ \mathbf{Y}_{\textrm{sps},[3] } $ involves a complexity of 
$ \mathcal{O}(K^3G^2N^4) $. 
Recovering delays using element-space ESPRIT introduces a complexity of 
$ \mathcal{O}(R^2KG^2) $. 
The complexity of constructing $ \{\hat{\mathbf{b}}_{n,r}\}_{r=1}^R, n=2,3,4,5 $ via SVD is in the order of 
$ \mathcal{O}( R(G^3+N^3) ) $. 
In CE Stage~I, the column-wise transformed-space ESPRIT takes the complexity of 
$ \mathcal{O}( CG+RN ) $. 
In Stage~II, we assume that $ E_n=E_{n'} $ and $ I_n=I_{n'} $ for distinct $ n $ and $ n' $, thus we ignore their subscript $ n $. 
The CBS involves a complexity of 
$ \mathcal{O}( IE(CG+RN) ) $. 
In Stage~III, the complexity of ALS is in the order of 
$ \mathcal{O}( TRKGN ( KGN + RGN + RKN + RKG )  ) $. 
In a nutshell, our complexities at the end of CE Stages~I\textendash III are 
\begin{align}
	\mathcal{O}_{\textrm{I}} &= 
	\mathcal{O}(K^3G^2N^4) + \mathcal{O}( R^2KG^2 ) + \mathcal{O}( RG^3 ) + \mathcal{O}( RN^3 ),
	\nonumber \\
	\mathcal{O}_{\textrm{II}} &= \mathcal{O}_{\textrm{I}} + 
	\mathcal{O}( IECG ) + \mathcal{O}( IERN ),
	\nonumber \\
	\mathcal{O}_{\textrm{III}} &= 
	\mathcal{O}(K^3G^2N^4) + \mathcal{O} (TRK^2G^2N^2) + \mathcal{O} (TR^2KG^2N^2) 
	\nonumber \\ &+ 
	\mathcal{O} (TR^2K^2GN^2) + \mathcal{O} (TR^2K^2G^2N) + \mathcal{O}( RG^3 )
	\nonumber \\ &+ 
	\mathcal{O}( RN^3 ) + \mathcal{O}( IECG ) + \mathcal{O}( IERN ). 
	\label{eq:my_comp}
\end{align}

The work in \cite{Zhou2017} introduced a pioneering tensor-based CE method. This approach initially applies ALS-CPD to extract the factors, followed by an exhaustive grid-based CBS to recover the channel parameters. The ALS-CPD's complexity is in the same order as described before \eqref{eq:my_comp}. Additionally, assuming that $S$  searches are required for each exhaustive CBS, the total complexity of the CBS process is $ \mathcal{O}( S( RK + CG+ RN) ) $. 
In conclusion, the overall complexity comes to 
\begin{align}
	\mathcal{O}& (TRK^2G^2N^2) + \mathcal{O} (TR^2KG^2N^2) + 
	\mathcal{O} (TR^2K^2GN^2)
	\nonumber \\ &+ 
	\mathcal{O} (TR^2K^2G^2N) + \mathcal{O}(SRK) + \mathcal{O}(SCG)+ \mathcal{O}(SRN).
	\nonumber
\end{align}
Note that the iteration number of ALS-CPD required here is larger than ours in Stage~III, as our Stage~II has generated an accurate initialization for the subsequent iterative ALS. 
Furthermore, to attain comparable search performance, CBS here also requires much more searches than ours in Stage~II, i.e., $ S > IE $, since we employ a modified iterative strategy and input coarse results obtained from Stage~I. 

Study \cite{Zheng2024} proposed a serial tensor-based CE approach tailored for the fully LOS scenario ($L=C=1,R=2$), a particular case of our multipath consideration. 
This mechanism sequentially involves CPD+ESPRIT two times and optional LS refinement. 
The complexities of CPDs are in the order of 
$ \mathcal{O}( TRN(KN+RN+RK) ) $ and 
$ \mathcal{O}( TCG(G+C) ) $. 
The ESPRIT estimations introduce complexities of 
$ \mathcal{O}( R(K+N ) $ and $ \mathcal{O}( CG ) $.  
However, the complexity of the LS step cannot be evaluated well in theory since this issue is smartly addressed by the MATLAB toolbox without a specific solution. 
The complexity at the end of the ESPRIT step (without LS refinement)  becomes 
\begin{equation}
	\mathcal{O}(TRKN^2) + \mathcal{O}(TR^2N^2) + \mathcal{O}(TR^2KN) + \mathcal{O}(TG^2). 
\end{equation}

\section{Numerical Results}\label{sec:NR}
This section evaluates the CE performance of our algorithms via simulation experiments. The derived CRLB in Section~\ref{sec:CRLB} is applied as an estimation accuracy metric.
Furthermore, benchmarking accuracy is provided from existing methods: CPD+ESPRIT+LS \cite{Zheng2024} (optional LS refinement), CPD+CBS \cite{Zhou2017}, and VSCPD+CBS \cite{He2024}. 
Notably, the last two baseline schemes above originally did not aim to solve the current CE issue in this paper. Unlike our modified iterative search strategy in CE Stage II, they have simply invoked exhaustive grid-based CBS. Their exhaustive search scheme is retained, but to apply to the current CE scenario, CPD is extended to decompose expanded tensor $ \hat{\bm{\mathcal{Y}}}_{\textrm{sps}} $ in CPD+CBS and directly employs our VSCPD in VSCPD+CBS. 
All experiments are conducted in a PC using MATLAB R2022b with Intel(R) Core(TM) i9-14900HX 2.20 GHz CPU and 16GB RAM.

\subsection{Simulation Settings and Performance Metrics}


The array size of RIS is $ M_y=M_z=15 $. 
As for the size of BS, we set $ \tilde{N}_y=\tilde{N}_z=10 $ and $ N_1=N_2=5 $. 
The bandwidth is $320$~MHz with central carrier frequency $ f_{\textrm{c}}=28 $~GHz. 
There are $ K_0=128 $ subcarriers, of which the first $ K=32 $ subcarriers are pilots. The number of time slots is  $ G=49 $.  
The transmit power is set to $ \mathbf{P}_{\textrm{T}} = 7 $~dBm, and the power of active RIS is restricted to $ \mathbf{P}_{\textrm{R}} = 1.76 $~dBm.  
The noise power spectral densities (PSDs) at both BS and RIS are $ -174 $~dBm/Hz, and their noise figures are identically $ 10 $~dB. 
The element spacings of BS and RIS are $ d_{\textrm{B}}=0.5\lambda $ and  $ d_{\textrm{R}}=0.1\lambda $, respectively, and wavelength $ \lambda=\frac{c}{f_{\textrm{c}}} $ with light speed $c$.  The multipath numbers are  $ L=P=Q=2 $, and thus $ R=6 $. 
Delays, AOAs, AODs, and path gains are generated according to practical geometric models specified in \cite{Zheng2024}, and the same LOS settings are employed as shown in \cite[Table~I]{Zheng2024}.
Additionally, the scatterers for UE-BS, UE-RIS, and RIS-BS links are located at 
$ [4, 4, 2]^T $~m, $ [-2, -4, 1]^T $~m, and $ [-1, 1, 5]^T $~m, respectively.  
The delays are expressed in meter units by multiplying light speed, and the angles are in degree units. 
The received signal-to-noise ratio (SNR) is defined as 
\begin{equation}
	\operatorname{SNR} = 
	\frac{ \sum_{k=1}^K \sum_{g=1}^G \|\mathbf{y}^{(k,g)}\|^2 }
	{ \sum_{k=1}^K \sum_{g=1}^G \operatorname{tr} (\mathbf{C}^{(k,g)})  }. 
\end{equation}

Tensor operations are programmed based on the Tensorlab toolbox \cite{Vervliet2016}.  
The spatial-smoothing parameter is set to $ K_1 = 15 $ with $ K_1+K_2-1=K $. 
Always, set  $ G_1=G_2 $ with $ G=G_1G_2 $. 
The generators of Vandermonde matrices $ \{\mathbf{T}_n\}_{n=2}^5 $ are randomly drawn.
In our CE Stage~II, $ E_n=2\tilde{E}+1 $ with $ \tilde{E}=100 $, $ I_n = 8 $, $ \zeta_n = 0.5$, and $ \Delta_n^1 = \frac{U_n}{10\tilde{C}} $ with parameter range $ (-U_n, U_n) $. 
To appraise the search accuracy using identical search numbers, we use $1608 $  searches in the CBS part of the VSCPD+CBS scheme.  
Moreover, the search number in CBS of CPD+CBS approach is set as $ 10^4 $, much greater than ours. 
In the proposed CE Stage~III, the convergence tolerance for ALS is set to $ 10^{-8} $. 
The convergence tolerance for ALS-CPD in benchmarking methods is designed as $ 10^{-15} $, which is significantly smaller than ours, as the ALS-CPD performance becomes undesirable using a larger tolerance in multipath scenarios. 
The LS refinement in the CPD+ESPRIT+LS approach is realized by Manopt toolbox \cite{Boumal2014} with $40$ times the maximum iterations. 

The preceding settings are applied to the following simulations unless otherwise specified.

Root mean squared error (RMSE) is used to evaluate the accuracy of channel-parameter estimators. That is,
\begin{equation}
	\operatorname{RMSE}\,(\bm{\xi}) = 
	\sqrt{ \mathbb{E}  \big[ \| \hat{\bm{\xi}} - \bm{\xi} \|^2 \big] }, 
\end{equation}
where $ \bm{\xi} $ can be any multipath-parameter vector/scalar, and $ \hat{\bm{\xi}} $ denotes its estimate. Due to the permutation ambiguity in the parameter estimates,  every RMSE is determined after sorting both the ground truth and estimated values. 
Furthermore, the estimation accuracy of the signal tensor is leveraged as a comprehensive performance metric for all channel-parameter estimates. The recovery precision of the signal tensor is evaluated using normalized MSE (NMSE), defined as 
\begin{equation}
	\operatorname{NMSE}\,( \bm{\mathcal{Y}} ) = 
	\frac{ \mathbb{E} \big[  \big\| \bm{\mathcal{Y}} ( \hat{\bm{\phi}}_{\textrm{int}} ) -
		\bm{\mathcal{Y}} (\bm{\phi}_{\textrm{int}}) \big\|_F^2  \big] }
	{ \| \bm{\mathcal{Y}} ( \bm{\phi}_{\textrm{int}} ) \|^2_F } ,  
\end{equation}
where $ \hat{\bm{\phi}}_{\textrm{int}} $ is the estimated version of $ \bm{\phi}_{\textrm{int}} $. 
The statistical expectation is approximately computed via the average result over $ 10^{3} $ Monte Carlo trials.

\begin{figure*}[t]
	\centering
	\includegraphics[width=0.86\linewidth]{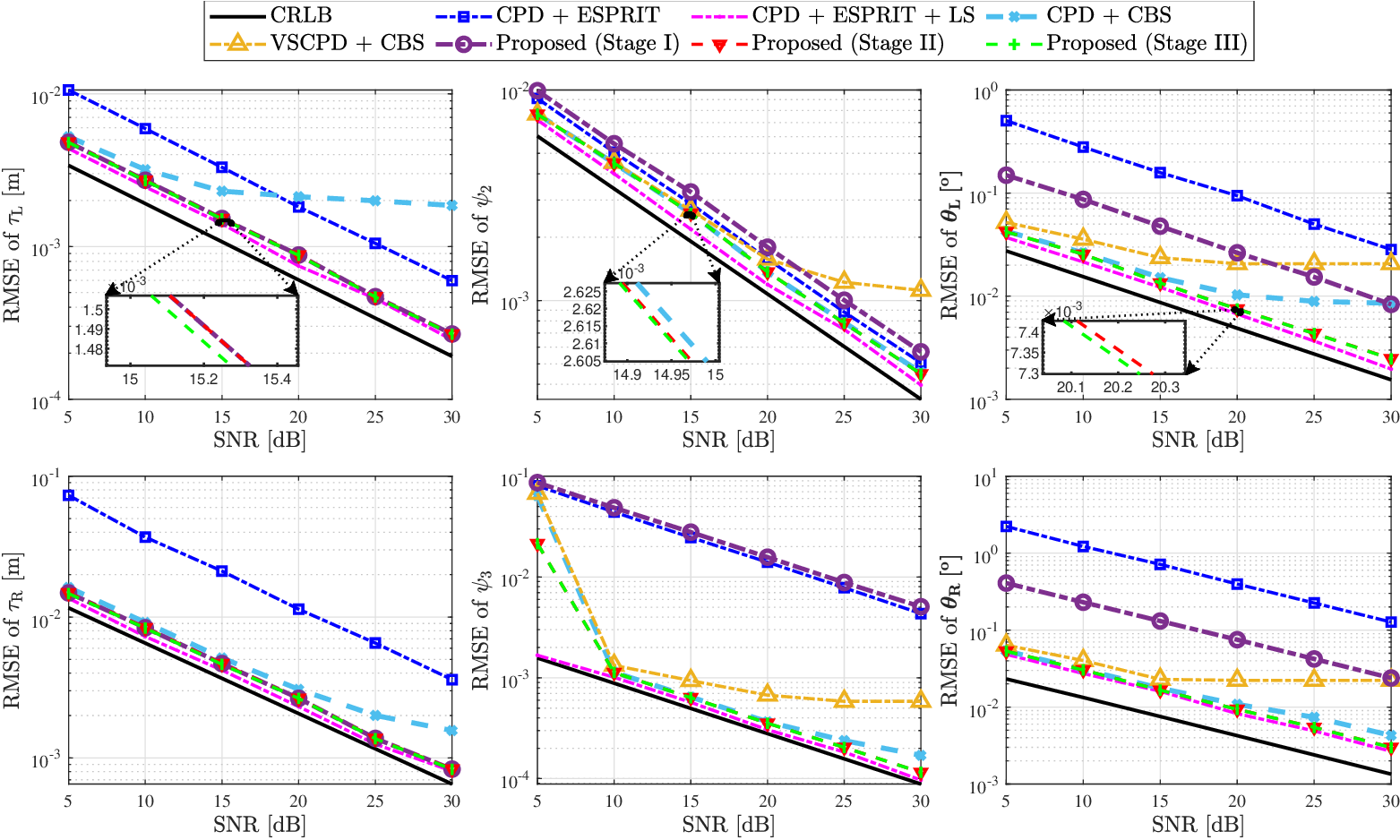}
	\caption{RMSEs of $ \tau_{\textrm{L}} $, $ \tau_{\textrm{R}} $, $ \psi_{2} $, $ \psi_{3} $, $ \bm{\theta}_{\textrm{L}} $, and $ \bm{\theta}_{\textrm{R}} $ versus SNR in LOS scenario.} 
\label{fig:RMSE_LOS}
\end{figure*}

\subsection{CE in LOS Scenario} 
Recall that the competing algorithm CPD+ESPRIT+LS is tailored for the LOS scenario, and its performance undoubtedly degrades in the multipath scenario. 
The estimation performance in the LOS case is first discussed for a fair comparison. 
For this,  remove all scatterers and let $ L=P=Q=1 $ with $ R=2 $. 
To reserve the parameter designs in the original CPD+ESPRIT+LS implementation, parameters are set as  $ G=9 $ and $ P_{\textrm{R}} = 7 $~dBm. 
In our Stage~III, the convergence tolerance for ALS is shrunk to $10^{-12}$ for better performance. 

\subsubsection{Estimation Accuracy} 
We range the SNR from $5$~dB to $30$~dB in $5$~dB increments. The estimation performance for channel parameters is depicted in Fig.~\ref{fig:RMSE_LOS}. The delay RMSEs from the VSCPD+CBS method are omitted, as the delay estimation is performed similarly. Algebraic ESPRIT-based methods generally show inferior performance, with significant gaps compared to the corresponding CRLBs. Fortunately, Stage~I significantly outperforms CPD+ESPRIT in delay and AOA accuracy thanks to our superior spatial smoothing-based VSCPD technique, which enhances tensor factorization performance across most modes compared to the ALS-CPD scheme.

We further see from Fig.~\ref{fig:RMSE_LOS} that 
CBS-based (or ML-like) methods can improve accuracy over ESPRIT methods. Stage~II demonstrates notable performance gains in high-SNR scenarios compared to CPD+CBS (which involves more extensive searches) and VSCPD+CBS (with the same number of searches), highlighting the effectiveness of our modified search strategy, where search space is iteratively constructed with coarse initialization achieved from Stage~I. Both Stage~III and the existing LS technique can further refine previous estimates, with LS refinement offering better performance. However, the accuracy of LS method costs significantly increased complexity (to be shown soon). Further, LS refinement is computationally impractical in multipath scenarios, where the number of optimization variables boosts remarkably.


\begin{figure}[t]
	\centering
	\includegraphics[width=1\linewidth]{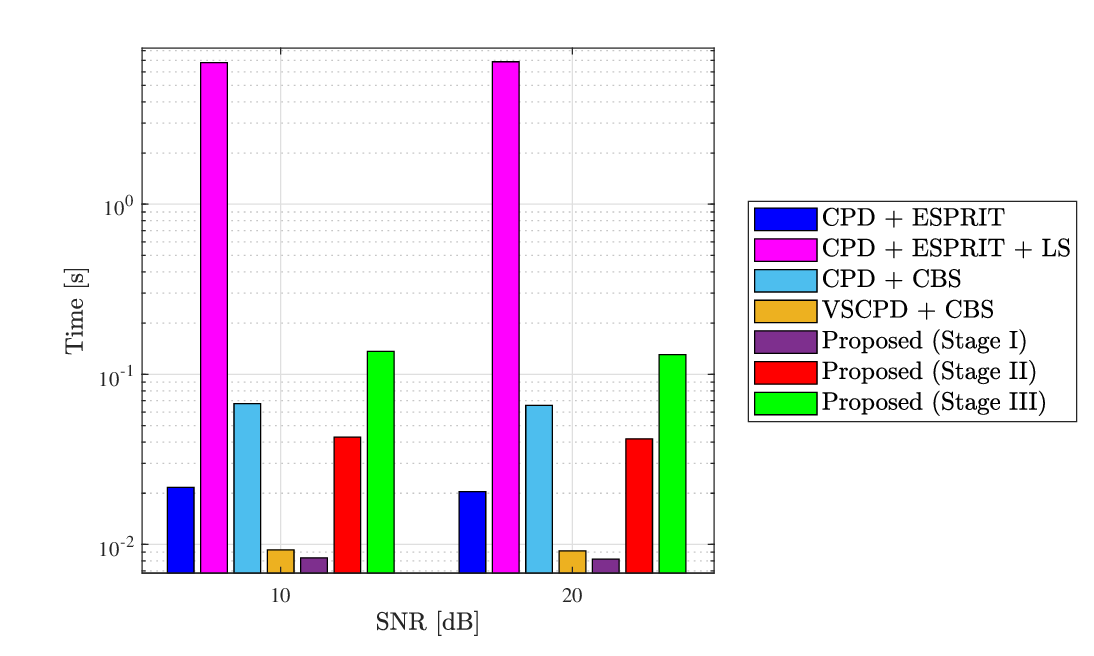}
	\caption{Average computation time with varying SNR in LOS scenario.} 
	\label{fig:complexity_LOS}
\end{figure}

\subsubsection{Implementation Complexity} 

The average CPU run time of different algorithms is presented in Fig.~\ref{fig:complexity_LOS}, with $ \operatorname{SNR}\in\{10,20\} $~dB. 
The computation time decreases marginally for each algorithm with ascending SNR. 
For algorithm comparison,  specific run time is listed hereafter for $10$~dB SNR. The complexity of our CE framework increases as the stage-wise process develops.  
Our Stage~I enjoys the lowest complexity ($8.35$~ms), only $38.5\%$ of CPD+ESPRIT ($21.67$~ms). 
This is owing to the algebraic implementation of our VSCPD in contrast to the iterative ALS-CPD. 
Furthermore, the complexity of our Stage II with a modified search strategy is lower than CPD+CBS and greater than VCPD+CBS. 
As mentioned, proposed Stage~II has an accuracy advantage over these two existing CBS-based methods, especially in high-SNR cases. 
Significantly, the complexity of our Stage~III ($0.14$~s) is only $2.1\%$ of LS-refined algorithm ($6.81$~s), tremendously saving the computation time with a theoretical specific solution.

\begin{figure*}[htbp]
	\centering
	\includegraphics[width=0.86\linewidth]{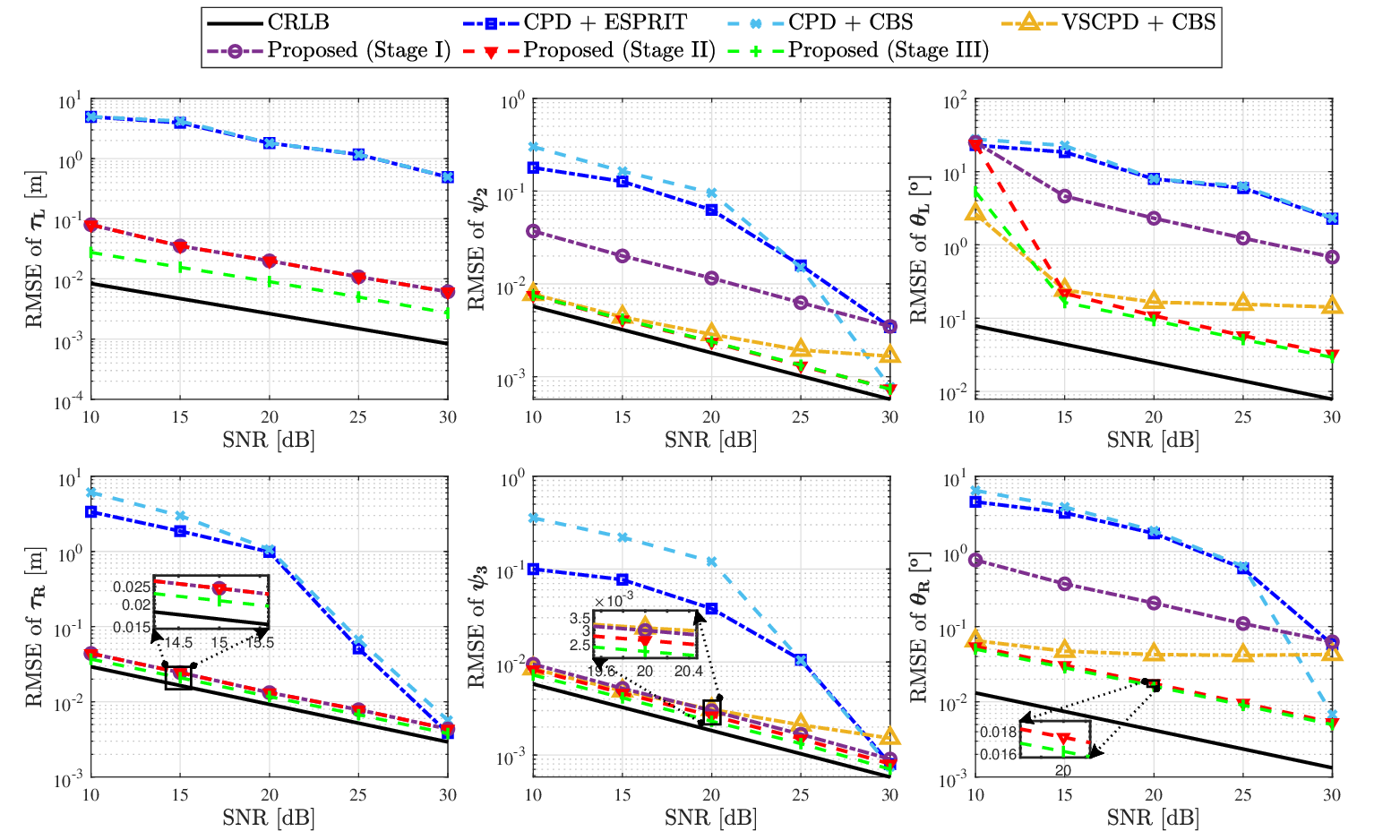}
	\caption{RMSEs of $ \bm{\tau}_{\textrm{L}} $, $ \bm{\tau}_{\textrm{R}} $, $ \bm{\psi}_{2} $, $ \bm{\psi}_{3} $, $ \bm{\theta}_{\textrm{L}} $, and $ \bm{\theta}_{\textrm{R}} $ versus SNR in multipath scenario.} 
	\label{fig:RMSE_MTP}
\end{figure*}

\begin{figure}[t]
	\centering
	\includegraphics[width=0.85\linewidth]{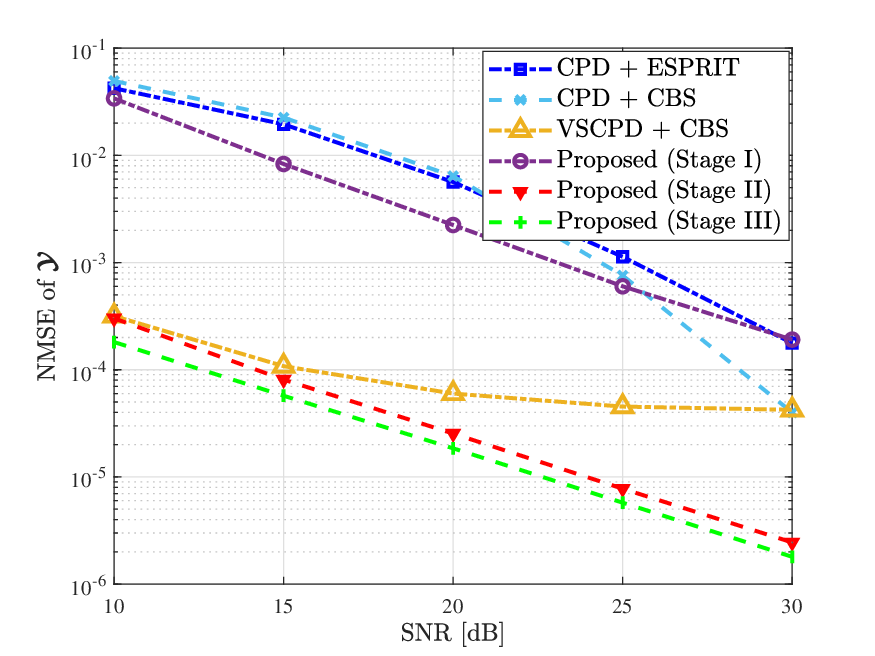}
	\caption{NMSE of $ \bm{\mathcal{Y}} $ versus SNR in multipath scenario.} 
	\label{fig:NMSE_SNR}
\end{figure}

\subsection{CE in Multipath Scenarios}
The CPD+ESPRIT method, initially designed for LOS scenarios, is extended to the multipath case to ensure the fair competition; however, the subsequent LS refinement is no longer feasible due to its significant computational complexity. Additionally, algorithms may fail based on the criteria outlined in Section~\ref{sec:CEI}, i.e., the {\it variance principle} and {\it similarity principle}. Each RMSE/NMSE result is reported from $10^3$ successful runs out of a minimum of $10^3$ total trials. In this context, the success rate is an additional evaluation metric for tensor decomposition performance, defined as the ratio of the number of successful runs (i.e., $10^3$) to the total number of trials.

\begin{table}[t]
	\centering
	\setlength{\tabcolsep}{8.2pt}
	\renewcommand\arraystretch{0.9}
	\caption{Success Rates of Tensor Decomposition versus SNR}
	\begin{tabular}{cccccc}
		\toprule
		& $10$~dB    & $15$~dB    & $20$~dB    & $25$~dB    & $30$~dB \\ 
		\midrule
		ALS-CPD & $83.4\%$ & $92.2\%$ & $97.3\%$ & $99.3\%$ & $99.8\%$ \\
		VSCPD  & $100\%$ & $100\%$ & $100\%$ & $100\%$ & $100\%$ \\ 
		\bottomrule
	\end{tabular}%
	\label{tab:rate_SNR}%
\end{table}%

\subsubsection{Estimation Accuracy in Varying-SNR Scenario}

In this scenario, the SNR varies from $10$ dB to $30$ dB in $5$ dB increments. 
Table~\ref{tab:rate_SNR} presents the success rates of the existing ALS-CPD and the proposed VSCPD in this varying-SNR scenario. While the probability of success for ALS-CPD increases with higher SNR, VSCPD keeps achieving perfect success, indicating its superior robustness against SNR. 
Moreover, the RMSEs of channel parameters and the NMSE of the signal tensor are presented in Fig.~\ref{fig:RMSE_MTP} and Fig.~\ref{fig:NMSE_SNR}, respectively. Delay results using VSCPD+CBS are omitted in Fig.~\ref{fig:RMSE_MTP}. The estimation performance of our CE framework improves sequentially through two stages of refinement, consistent with the results in Fig.~\ref{fig:RMSE_LOS}. The CPD+ESPRIT and CPD+CBS schemes exhibit similar yet poor accuracy, both inferior to Stage~I in general. This indicates that classical ALS-CPD is unsuitable for scenarios with rank-deficient factor matrices, and our VSCPD effectively addresses this deficiency issue by leveraging the Vandermonde structure. Furthermore, Stage~II outperforms VSCPD+CBS in accuracy under high-SNR conditions, benefiting from a superior iterative search scheme.\footnote{Observe from Fig.~\ref{fig:RMSE_MTP} that the RMSE of AOA $ \bm{\theta}_{\textrm{L}} $ from all algorithms leaves the CRLB far away. This is because CRLB is a local bound, providing tight estimation performance evaluation only in the asymptotic (i.e.,  high-SNR) region. In contrast, Ziv-Zakai bound (ZZB) \cite{Chen2024,Zhang2024c,Zhang2023a} keeps effective outside the asymptotic region and can serve as a global bound for the estimation accuracy from the low to high SNR situation. }

\begin{figure}[t]
	\centering
	\includegraphics[width=0.85\linewidth]{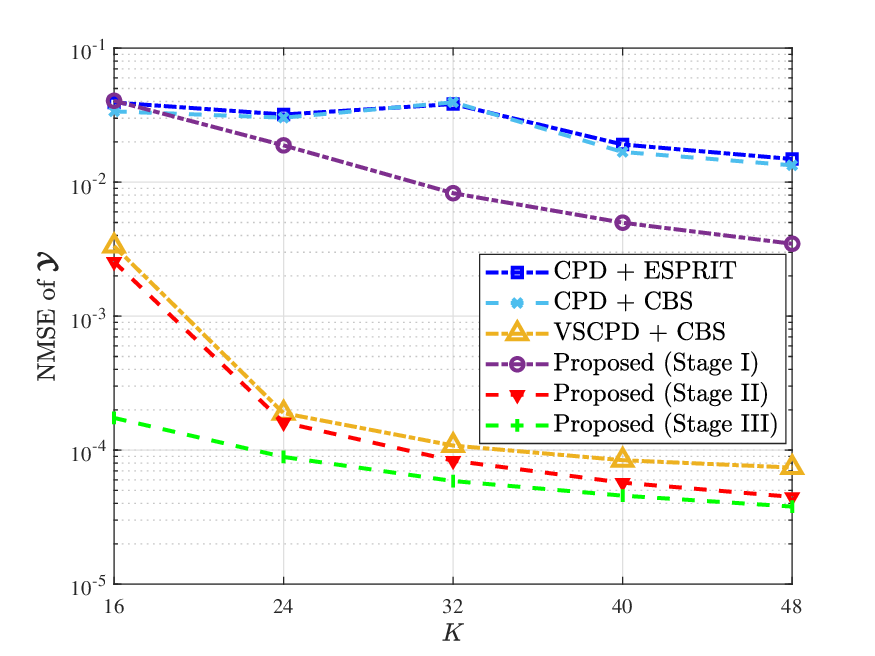}
	\caption{NMSE of $ \bm{\mathcal{Y}} $ versus $ K $ in multipath scenario.} 
	\label{fig:NMSE_K}
\end{figure}

\begin{figure}[t]
	\centering
	\includegraphics[width=1\linewidth]{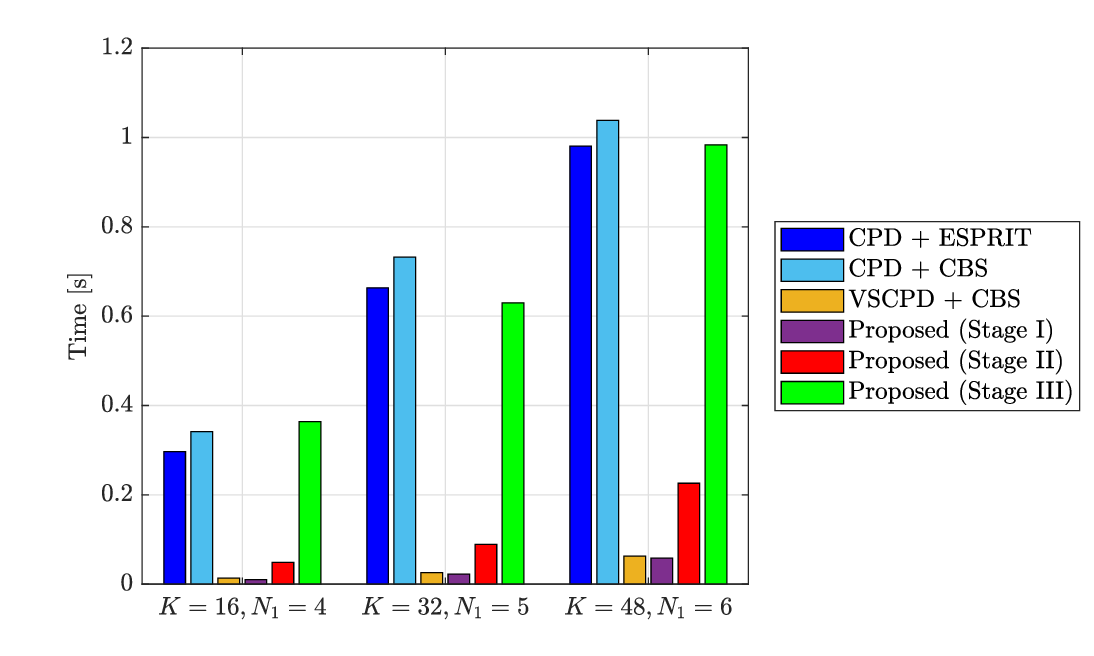}
	\caption{Average computation time with varying $K$ and $N_1=N_2$ in multipath scenario.} 
	\label{fig:complexity_MTP}
\end{figure}

\begin{table}[t]
	\centering
	\renewcommand\arraystretch{0.9}
	\caption{Success Rates of Tensor Decomposition versus $ K $}
	\begin{tabular}{cccccc}
		\toprule
		& $K=16$    & $K=24$    & $K=32$    & $K=40$   & $K=48$ \\ 
		\midrule
		ALS-CPD & $96.3\%$ & $95.5\%$ & $91.7\%$ & $91.5\%$ & $91.7\%$ \\
		VSCPD  & $100\%$ & $100\%$ & $100\%$ & $100\%$ & $100\%$ \\
		\bottomrule
	\end{tabular}%
	\label{tab:rate_K}%
\end{table}%

\begin{table}[ht]
	\centering
	\setlength{\tabcolsep}{2pt} 
	\renewcommand\arraystretch{0.9}
	\caption{Comparison between Active-RIS and Passive-RIS Cases}
	\begin{tabular}{ccccc}
		\toprule
		& Amp. Coef.  & Power Ratio & Success Rate & Tensor NMSE  \\
		\midrule
		Passive-RIS Case & $1$     & $8.3\times10^{-5}$ & $39.6\%$  & $61.9\times10^{-6}$ \\
		Active-RIS Case  & $103.3$ & $0.9$ & $100\%$  & $2.0\times10^{-6}$ \\
		\bottomrule
	\end{tabular}%
	\label{tab:2modes_com}%
\end{table}%

\subsubsection{Estimation Accuracy in Varying-$K$ Scenario} 

In this scenario, the number $K$ of training subcarrier is adjusted from $16$ to $48$ in increments of $8$, with $K_1$ consistently set to $0.5K$. The SNR remains fixed at $15$ dB. 
In this varying-$K$ scenario, the success rates of tensor decomposition are listed in Table~\ref{tab:rate_K},  and the NMSE results of tensor reconstruction are shown in Fig.~\ref{fig:NMSE_K}.  

We see that the success rate of ALS-CPD initially falls as $K$ grows and then keeps virtually the same when $K\ge32$. 
Moreover, the NMSEs of the two ALS-CPD-based approaches decrease with the growing $K$ in general. 
This is because that the growth of $K$ increases the tensor dimension, making CPD more challenging; while it provides more training subcarriers, benefiting estimator performance. 
We also witness at $K=32$ an abrupt increase of the ALS-CPD NMSEs, corresponding to the sharp decline in terms of the ALS-CPD success rate. 
By comparison, our VSCPD algorithm consistently achieves perfect success-rate results, demonstrating its significant robustness to the tensor dimension. 
Furthermore, the proposed approaches output superior NMSE performances in contrast to the ALS-CPD-based methods. 
The successive refinements within our CE framework are also evident in Fig.~\ref{fig:NMSE_K}.

\subsubsection{Implementation Complexity} 
The CPU run time for the multipath scenario is presented in Fig.~\ref{fig:complexity_MTP}. For the analysis, the parameter  $ K\in\{16,32,48\} $ with $ K_1=0.5K $ and $ N_1=N_2\in\{4,5,6\} $.  The SNR is fixed at $15$~dB. The results indicate that the complexity of each method increases with the rise in $K$ and $N_1$. Run time is provided for the case $ K=32, N_1=5 $ for a more detailed algorithm comparison.
 The Stage~I algorithm exhibits the lowest complexity at $22.26$~ms, slightly lower than that of the VSCPD+CBS method, which takes $25.97$~ms. Stage~II has an intermediate complexity of $88.97$~ms. Notably, Stage~III shows a complexity of $0.63$~s, comparable to those of two ALS-CPD approaches, highlighting the time-intensive nature of ALS-CPD for the multipath case.

\subsubsection{Active RIS versus Passive RIS}\label{sec:2modes_com}

As mentioned before \eqref{eq:PR}, the active RIS enables signal amplification while introducing extra noise. To gain insights into the comprehensive effect of the active RIS, we compare our active-RIS case with the conventional passive-RIS one and the comparison results are listed in Table~\ref{tab:2modes_com}. 
For the fair competition, we perform the simulations in both cases under $30$~dB SNR. Moreover, in the active-RIS case, we set the transmit power $P_{\textrm{T}}=7$~dBm and RIS power $P_{\textrm{R}}=0.1$~dBm; in the passive-RIS case, the transmit power is set as $ P_{\textrm{T}}+P_{\textrm{R}} $. 
The RIS amplification coefficient $ \eta $ is abbreviated as amp. coef. 
The power ratio is defined in the noise-free case as the ratio of signal power for cascaded link to the signal power for direct link.  
The success rate of tensor decomposition is obtained from our VSCPD. The NMSE of signal tensor estimation is acquired from the proposed CE Stage~II, since we find that our Stage~III cannot refine the accuracy over Stage~II in the passive case.  

We see from Table~\ref{tab:2modes_com} that the power of the cascaded channel can be significantly ameliorated through the amplification of RIS elements. Accordingly, both the success rate and the tensor NMSE improve remarkably compared with the passive-RIS case, although more noise is incurred from the active RIS. The reason is that the cascaded channel is too weak to recover accurately in the passive-RIS case, and hence the performances of tensor factorization and tensor reconstruction severely deteriorate.

\section{Conclusion}\label{sec:conc} 

This paper addressed the CE challenge for active RIS-assisted SIMO-OFDM wireless systems in fully multipath environments. Channels were modeled with multipath parameters, including path gains, delays, AOAs, and AODs, transforming the CE problem into a parameter recovery task. A fifth-order CP tensor was constructed from the received signal using customized designs. Involving five factor matrices, four of which suffered from rank deficiency, while the other exhibited a Vandermonde structure. Spatial smoothing was applied to extend the original tensor to the sixth-order one, and an efficient VSCPD method was developed to extract the factor matrices. A triple-stage CE algorithm, consisting of coarse estimation in Stage I and successive optional refinements in Stages II and III, was proposed based on 1D optimization. The CRLB was derived with the noise covariance matrix dependent on multipath parameters, and the implementation complexity of the proposed and benchmark algorithms was analyzed. Simulation results demonstrated the necessity of the amplification of active RIS elements and also the superiority of the proposed methods across various evaluation metrics. 

Future research directions include: (i) extension of current CE framework to the multiple-input multiple-output (MIMO) scenarios, (ii) the joint RIS calibration and user positioning problem using channel parameters, and (iii) the derivation of related globally tight ZZB.

\appendices

\renewcommand\thesubsectiondis{\thesection.\arabic {subsection}}

\section{Partial Derivatives in FIM calculation}\label{sec:Derivatives}

Firstly, we define 
\begin{gather}
	\beta_{ \textrm{L},\mathfrak{R}}^{(\ell)} = \mathfrak{R}\big( \beta_{\textrm{L}}^{(\ell)} \big), \;
	\beta_{\textrm{L},\mathfrak{I}}^{(\ell)} = \mathfrak{I}\big( \beta_{\textrm{L}}^{(\ell)} \big),
	\nonumber \\
	\beta_{\textrm{R},\mathfrak{R}}^{(p,q)} = \mathfrak{R}\big( \beta_{\textrm{R}}^{(p,q)} \big), \;
	\beta_{\textrm{R},\mathfrak{I}}^{(p,q)} = \mathfrak{I}\big( \beta_{\textrm{R}}^{(p,q)} \big),
	\nonumber \\
	\varOmega^{(k)} = 2 \pi f^{(k)}, 
	\nonumber \\
	\delta_{\textrm{L}}^{(k,\ell)} = e^{-\jmath \varOmega^{(k)} \tau_{\textrm{L}}^{(\ell)}}, \;
	\delta_{\textrm{R}}^{(k,p,q)} = e^{-\jmath \varOmega^{(k)} \tau_{\textrm{R}}^{(p,q)}},
	\nonumber \\
	\dot{\delta}_{\textrm{L}}^{(k,\ell)} = -\jmath \varOmega^{(k)} \delta_{\textrm{L}}^{(k,\ell)}, \;
	\dot{\delta}_{\textrm{R}}^{(k,p,q)} = -\jmath \varOmega^{(k)} \delta_{\textrm{R}}^{(k,p,q)},
	\nonumber \\
	\breve{\mathbf{a}}_{\textrm{R}}^{(p,q)} = 
	\mathbf{a}_{\textrm{R}} \big( \bm{\varphi}_{\textrm{A}}^{(p)} \big) \circledast
	\mathbf{a}_{\textrm{R}} \big( \bm{\varphi}_{\textrm{D}}^{(q)} \big),
	\nonumber \\
	\dot{\breve{\mathbf{a}}}_{\textrm{R},2(3)}^{(p,q)} = \jmath\frac{2\pi}{\lambda}
	\breve{\mathbf{a}}_{\textrm{R}}^{(p,q)} \circledast [\mathbf{P}_{\textrm{R}}]_{:,2(3)},
	\nonumber \\
	\bm{\theta} = [ \theta_{\textrm{az}}, \theta_{\textrm{el}} ]^T, 
	\nonumber \\
	\dot{\mathbf{d}}_{\textrm{az}}(\bm{\theta}) = 
	\big[ 0, \cos(\theta_{\textrm{az}}) \cos(\theta_{\textrm{el}}), 0 \big]^T, 
	\nonumber \\
	\dot{\mathbf{d}}_{\textrm{el}}(\bm{\theta}) = 
	\big[ 0, -\sin(\theta_{\textrm{az}}) \sin(\theta_{\textrm{el}}), \cos(\theta_{\textrm{el}}) \big]^T, 
	\nonumber \\
	\dot{\mathbf{a}}_{\textrm{B,az(el)}} (\bm{\theta}) = \jmath\frac{2\pi}{\lambda}
	\mathbf{a}_{\textrm{B}} (\bm{\theta}) \circledast 
	\big( \mathbf{P}_{\textrm{B}} \dot{\mathbf{d}}_{\textrm{az(el)}}(\bm{\theta})  \big), 
	\nonumber \\
	\mathbf{p}_{\textrm{B},i,j} =
	[\mathbf{P}_{\textrm{B}}]_{i,:}^T- [\mathbf{P}_{\textrm{B}}]_{j,:}^T, 
	\nonumber \\
	[\dot{\mathbf{A}}_{\textrm{B,az(el)}} (\bm{\theta})]_{i,j} = \jmath \frac{2\pi}{\lambda}
	\big( \mathbf{p}_{\textrm{B},i,j}^T \dot{\mathbf{d}}_{\textrm{az(el)}}(\bm{\theta}) \big)
	[\mathbf{A}_{\textrm{B}} (\bm{\theta})]_{i,j},\; i\neq j, 
	\nonumber \\
	[\dot{\mathbf{A}}_{\textrm{B,az(el)}} (\bm{\theta})]_{i,j} = 0,\; i= j . 
	\nonumber 
\end{gather}

Note that for complex-valued function $ \mathbf{F}(\phi) $ with respect to real-valued scalar $ \phi $, we have 
\begin{equation} 
	\frac{ \partial \mathfrak{R} \big(\mathbf{F}(\phi)\big) } { \partial\phi } = 
	\mathfrak{R} \left(  \frac{ \partial\mathbf{F}(\phi) } { \partial\phi } \right),   \;
	\frac{ \partial \mathfrak{I} \big(\mathbf{F}(\phi)\big) } { \partial\phi } = 
	\mathfrak{I} \left(  \frac{ \partial\mathbf{F}(\phi) } { \partial\phi } \right). 
	\nonumber 
\end{equation}
Furthermore, notice that $ \mathbf{y}_{\textrm{L}}^{(k,g)} $ degrades into $ \mathbf{y}_{\textrm{L}}^{(k)} $ in our case due to the identical pilot design before \eqref{eq:x}.  

In the following, we only list the non-zero partial derivatives. 
Other derivatives required by the FIM computation are all zeros.

\subsection{Derivatives in the First Term of \eqref{eq:FIM}}
\begin{align}
	\frac{ \partial \mathbf{y}_{\textrm{L}}^{(k)} }{ \partial \tau_{\textrm{L}}^{(\ell)} } &= 
	x \beta_{\textrm{L}}^{(\ell)} 
	\dot{\delta}_{\textrm{L}}^{(k,\ell)} 
	\mathbf{R}^H \mathbf{a}_{\textrm{B}} \big( \bm{\theta}_{\textrm{L}}^{(\ell)} \big),
	\nonumber \\
	\frac{ \partial \mathbf{y}_{\textrm{R}}^{(k,g)} }{ \partial \tau_{\textrm{R}}^{(p,q)} } &= 
	x \beta_{\textrm{R}}^{(p,q)} \rho_{\textrm{R}}^{(p,q)} 
	\dot{\delta}_{\textrm{R}}^{(k,p,q)} 
	\mathbf{R}^H \mathbf{a}_{\textrm{B}} \big( \bm{\theta}_{\textrm{R}}^{(q)} \big), 
	\nonumber \\
	\frac{ \partial \mathbf{y}_{\textrm{R}}^{(k,g)} }{ \partial \psi_{2(3)}^{(p,q)} } &= 
	x \beta_{\textrm{R}}^{(p,q)} \delta_{\textrm{R}}^{(k,p,q)} 	
	\big( (\bm{\gamma}^{(g)})^T	\dot{\breve{\mathbf{a}}}_{\textrm{R},2(3)}^{(p,q)} \big)
	\mathbf{R}^H \mathbf{a}_{\textrm{B}} \big( \bm{\theta}_{\textrm{R}}^{(q)} \big),
	\nonumber \\
	\frac{ \partial \mathbf{y}_{\textrm{L}}^{(k)} }{ \partial \theta_{\textrm{L,az(el)}}^{(\ell)} } &= 
	x
	\beta_{\textrm{L}}^{(\ell)} \delta_{\textrm{L}}^{(k,\ell)} \mathbf{R}^H  
	\dot{\mathbf{a}}_{\textrm{B,az(el)}} \big( \bm{\theta}_{\textrm{L}}^{(\ell)} \big),
	\nonumber \\
	\frac{ \partial \mathbf{y}_{\textrm{R}}^{(k,g)} }{ \partial \theta_{\textrm{R,az(el)}}^{(q)} } &= 
	\sum_{p=1}^P
	x 
	\beta_{\textrm{R}}^{(p,q)} \rho_{\textrm{R}}^{(p,q)} \delta_{\textrm{R}}^{(k,p,q)} \mathbf{R}^H  
	\dot{\mathbf{a}}_{\textrm{B,az(el)}} \big( \bm{\theta}_{\textrm{R}}^{(q)} \big),   
	\nonumber \\	 
	\frac{ \partial \mathbf{y}_{\textrm{L}}^{(k)} }
	{ \partial \beta_{\textrm{L},\mathfrak{R}}^{(\ell)} } &= 
	x \delta_{\textrm{L}}^{(k,\ell)} 
	\mathbf{R}^H \mathbf{a}_{\textrm{B}} \big( \bm{\theta}_{\textrm{L}}^{(\ell)} \big),
	\;
	\frac{ \partial \mathbf{y}_{\textrm{L}}^{(k)} }
	{ \partial \beta_{\textrm{L},\mathfrak{I}}^{(\ell)} } =
	j \frac{ \partial \mathbf{y}_{\textrm{L}}^{(k)} }
	{ \partial \beta_{\textrm{L},\mathfrak{R}}^{(\ell)} },
	\nonumber \\
	\frac{ \partial \mathbf{y}_{\textrm{R}}^{(k,g)} }
	{ \partial \beta_{\textrm{R},\mathfrak{R}}^{(p,q)} } &= 
	x \rho_{\textrm{R}}^{(p,q)} \delta_{\textrm{R}}^{(k,p,q)} 
	\mathbf{R}^H \mathbf{a}_{\textrm{B}} \big( \bm{\theta}_{\textrm{R}}^{(q)} \big),
	\,
	\frac{ \partial \mathbf{y}_{\textrm{R}}^{(k,g)} }
	{ \partial \beta_{\textrm{R},\mathfrak{I}}^{(p,q)} } =
	j \frac{ \partial \mathbf{y}_{\textrm{R}}^{(k,g)} }
	{ \partial \beta_{\textrm{R},\mathfrak{R}}^{(p,q)} }.
	\nonumber 
\end{align}

\subsection{Derivatives in the Second Term of \eqref{eq:FIM}}

\begin{align}
	\frac{ \partial \mathbf{G}_{\textrm{R}}\mathbf{G}_{\textrm{R}}^H }
	{ \partial \theta_{\textrm{R,az(el)}}^{(q)} } &= 
	\eta^2 M_y M_z \big|\beta_{\textrm{R},2}^{(q)}\big|^2 
	\mathbf{R}^H 
	\dot{\mathbf{A}}_{\textrm{B,az(el)}} \big( \bm{\theta}_{\textrm{R}}^{(q)} \big)
	\mathbf{R}, 
	\nonumber \\
	\frac{ \partial \mathbf{G}_{\textrm{R}}\mathbf{G}_{\textrm{R}}^H }
	{ \partial \big|\beta_{\textrm{R},2}^{(q)}\big|^2 } &= 
	\eta^2 M_y M_z 
	\mathbf{R}^H 
	\mathbf{A}_{\textrm{B}} \big( \bm{\theta}_{\textrm{R}}^{(q)} \big)
	\mathbf{R}. 
	\nonumber
\end{align}

\bibliographystyle{IEEEtran}
\bibliography{IEEEabrv,Ref.bib}

\begin{thebibliography}{10}
\providecommand{\url}[1]{#1}
\csname url@samestyle\endcsname
\providecommand{\newblock}{\relax}
\providecommand{\bibinfo}[2]{#2}
\providecommand{\BIBentrySTDinterwordspacing}{\spaceskip=0pt\relax}
\providecommand{\BIBentryALTinterwordstretchfactor}{4}
\providecommand{\BIBentryALTinterwordspacing}{\spaceskip=\fontdimen2\font plus
\BIBentryALTinterwordstretchfactor\fontdimen3\font minus
  \fontdimen4\font\relax}
\providecommand{\BIBforeignlanguage}[2]{{%
\expandafter\ifx\csname l@#1\endcsname\relax
\typeout{** WARNING: IEEEtran.bst: No hyphenation pattern has been}%
\typeout{** loaded for the language `#1'. Using the pattern for}%
\typeout{** the default language instead.}%
\else
\language=\csname l@#1\endcsname
\fi
#2}}
\providecommand{\BIBdecl}{\relax}
\BIBdecl

\bibitem{Wu2019a}
Q.~Wu and R.~Zhang, ``Intelligent reflecting surface enhanced wireless network
  via joint active and passive beamforming,'' \emph{IEEE Trans. Wireless
  Commun.}, vol.~18, no.~11, pp. 5394--5409, Nov. 2019.

\bibitem{Li2022d}
R.~Li, B.~Guo, M.~Tao, Y.-F. Liu, and W.~Yu, ``Joint design of hybrid
  beamforming and reflection coefficients in {RIS}-aided {mmWave MIMO}
  systems,'' \emph{IEEE Trans. Commun.}, vol.~70, no.~4, pp. 2404--2416, Apr.
  2022.

\bibitem{Wu2021}
Q.~Wu, S.~Zhang, B.~Zheng, C.~You, and R.~Zhang, ``Intelligent reflecting
  surface-aided wireless communications: A tutorial,'' \emph{IEEE Trans.
  Commun.}, vol.~69, no.~5, pp. 3313--3351, May 2021.

\bibitem{Huang2019}
C.~Huang, A.~Zappone, G.~C. Alexandropoulos, M.~Debbah, and C.~Yuen,
  ``Reconfigurable intelligent surfaces for energy efficiency in wireless
  communication,'' \emph{IEEE Trans. Wireless Commun.}, vol.~18, no.~8, pp.
  4157--4170, Aug. 2019.

\bibitem{Liu2024b}
L.~Liu, S.~Zhang, and S.~Cui, ``Leveraging a variety of anchors in cellular
  network for ubiquitous sensing,'' \emph{IEEE Commun. Mag.}, vol.~62, no.~9,
  pp. 98--104, Sep. 2024.

\bibitem{Liu2021}
{Y. Liu et al.}, ``Reconfigurable intelligent surfaces: Principles and
  opportunities,'' \emph{IEEE Commun. Surv. \& Tut.}, vol.~23, no.~3, pp.
  1546--1577, Aug. 2021.

\bibitem{Long2021}
R.~Long, Y.-C. Liang, Y.~Pei, and E.~G. Larsson, ``Active reconfigurable
  intelligent surface-aided wireless communications,'' \emph{IEEE Trans.
  Wireless Commun.}, vol.~20, no.~8, pp. 4962--4975, Aug. 2021.

\bibitem{Schroeder2022}
R.~Schroeder, J.~He, G.~Brante, and M.~Juntti, ``Two-stage channel estimation
  for hybrid {RIS} assisted {MIMO} systems,'' \emph{IEEE Trans. Commun.},
  vol.~70, no.~7, pp. 4793--4806, Jul. 2022.

\bibitem{Stuber2004}
G.~Stuber, J.~Barry, S.~McLaughlin, Y.~Li, M.~Ingram, and T.~Pratt, ``Broadband
  {MIMO-OFDM} wireless communications,'' \emph{Proc. IEEE}, vol.~92, pp.
  271--294, Feb. 2004.

\bibitem{Barhumi2003}
I.~Barhumi, G.~Leus, and M.~Moonen, ``Optimal training design for {MIMO OFDM}
  systems in mobile wireless channels,'' \emph{IEEE Trans. Signal Process.},
  vol.~51, pp. 1615--1624, Jun. 2003.

\bibitem{Chen2021}
H.~Chen, F.~Ahmad, S.~Vorobyov, and F.~Porikli, ``Tensor decompositions in
  wireless communications and {MIMO} radar,'' \emph{IEEE J. Sel. Topics Signal
  Process.}, vol.~15, no.~3, pp. 438--453, Apr. 2021.

\bibitem{Gong2023}
X.~Gong, W.~Chen, L.~Sun, J.~Chen, and B.~Ai, ``An {ESPRIT}-based supervised
  channel estimation method using tensor train decomposition for {mmWave 3-D
  MIMO-OFDM} systems,'' \emph{IEEE Trans. Signal Process.}, vol.~71, pp.
  555--570, 2023.

\bibitem{Zhou2016}
{Z. Zhou et al.}, ``Channel estimation for millimeter-wave multiuser {MIMO}
  systems via {PARAFAC} decomposition,'' \emph{IEEE Trans. Wireless Commun.},
  vol.~15, no.~11, pp. 7501--7516, 2016.

\bibitem{Wei2021}
{L. Wei et al.}, ``Channel estimation for {RIS}-empowered multi-user {MISO}
  wireless communications,'' \emph{IEEE Trans. Commun.}, vol.~69, no.~6, pp.
  4144--4157, Jun. 2021.

\bibitem{Araujo2021}
G.~T. de~Araujo, A.~L.~F. de~Almeida, and R.~Boyer, ``Channel estimation for
  intelligent reflecting surface assisted {MIMO} systems: A tensor modeling
  approach,'' \emph{IEEE J. Sel. Topics Signal Process.}, vol.~15, no.~3, pp.
  789--802, Apr. 2021.

\bibitem{Du2023}
J.~Du, Y.~Cheng, L.~Jin, and F.~Gao, ``Time-varying phase noise estimation,
  channel estimation, and data detection in {RIS}-assisted {MIMO} systems via
  tensor analysis,'' \emph{IEEE Trans. Signal Process.}, vol.~71, pp.
  3426--3441, 2023.

\bibitem{Shi2022}
X.~Shi, J.~Wang, and J.~Song, ``Triple-structured compressive sensing-based
  channel estimation for {RIS}-aided {MU-MIMO} systems,'' \emph{IEEE Trans.
  Wireless Commun.}, vol.~21, no.~12, pp. 11\,095--11\,109, Dec. 2022.

\bibitem{Zhao2022a}
S.~Zhao, N.~Guo, X.-P. Zhang, X.~Cui, and M.~Lu, ``Sequential
  {Doppler}-shift-based optimal localization and synchronization with {TOA},''
  \emph{IEEE Internet Things J.}, vol.~9, pp. 16\,234--16\,246, Sep. 2022.

\bibitem{Huang2023}
{Y. Huang et al.}, ``Joint localization and environment sensing by harnessing
  {NLOS} components in {RIS}-aided {mmWave} communication systems,'' \emph{IEEE
  Trans. Wireless Commun.}, vol.~22, no.~12, pp. 8797--8813, Dec. 2023.

\bibitem{Lin2021a}
Y.~Lin, S.~Jin, M.~Matthaiou, and X.~You, ``Tensor-based algebraic channel
  estimation for hybrid {IRS}-assisted {MIMO-OFDM},'' \emph{IEEE Trans.
  Wireless Commun.}, vol.~20, no.~6, pp. 3770--3784, Jun. 2021.

\bibitem{Xu2022}
X.~Xu, S.~Zhang, F.~Gao, and J.~Wang, ``Sparse bayesian learning based channel
  extrapolation for {RIS} assisted {MIMO-OFDM},'' \emph{IEEE Trans. Commun.},
  vol.~70, no.~8, pp. 5498--5513, Aug. 2022.

\bibitem{Mo2023}
L.~Mo, F.~Saggese, X.~Lu, Z.~Wang, and P.~Popovski, ``Direct tensor-based
  estimation of broadband {mmWave} channels with {RIS},'' \emph{IEEE Commun.
  Lett.}, vol.~27, no.~7, pp. 1849--1853, Jul. 2023.

\bibitem{Lin2022}
Y.~Lin, S.~Jin, M.~Matthaiou, and X.~You, ``Channel estimation and user
  localization for {IRS}-assisted {MIMO-OFDM} systems,'' \emph{IEEE Trans.
  Wireless Commun.}, vol.~21, pp. 2320--2335, Apr. 2022.

\bibitem{Zheng2022}
X.~Zheng, P.~Wang, J.~Fang, and H.~Li, ``Compressed channel estimation for
  {IRS}-assisted millimeter wave {OFDM} systems: A low-rank tensor
  decomposition-based approach,'' \emph{IEEE Wireless Commun. Lett.}, vol.~11,
  no.~6, pp. 1258--1262, Jun. 2022.

\bibitem{He2024}
F.~He, A.~Harms, and L.~Y. Yang, ``Slow-moving channel estimation via
  {Vandermonde} structured tensor decomposition in {RIS}-aided {MIMO}
  systems,'' \emph{IEEE Access}, vol.~12, pp. 67\,772--67\,783, 2024.

\bibitem{Sorensen2013}
M.~Sorensen and L.~De~Lathauwer, ``Blind signal separation via tensor
  decomposition with {Vandermonde} factor: {Canonical} polyadic
  decomposition,'' \emph{IEEE Trans. Signal Process.}, vol.~61, no.~22, pp.
  5507--5519, Nov. 2013.

\bibitem{Zhang2024}
{R. Zhang et al.}, ``Integrated sensing and communication with massive {MIMO}:
  A unified tensor approach for channel and target parameter estimation,''
  \emph{IEEE Trans. Wireless Commun.}, vol.~23, no.~8, pp. 8571--8587, Aug.
  2024.

\bibitem{Zhang2022a}
{R. Zhang et al.}, ``Tensor decomposition-based channel estimation for hybrid
  {mmWave} massive {MIMO} in high-mobility scenarios,'' \emph{IEEE Trans.
  Commun.}, vol.~70, no.~9, pp. 6325--6340, Sep. 2022.

\bibitem{Zhang2024a}
{R. Zhang et al.}, ``Channel training-aided target sensing for {Terahertz}
  integrated sensing and massive {MIMO} communications,'' \emph{IEEE Internet
  Things J.}, {Early Access,} 2024.

\bibitem{Wei2021a}
X.~Wei, D.~Shen, and L.~Dai, ``Channel estimation for {RIS} assisted wireless
  communications—part {I}: Fundamentals, solutions, and future
  opportunities,'' \emph{IEEE Commun. Lett.}, vol.~25, no.~5, pp. 1398--1402,
  May 2021.

\bibitem{He2020}
Z.-Q. He and X.~Yuan, ``Cascaded channel estimation for large intelligent
  metasurface assisted massive {MIMO},'' \emph{IEEE Wireless Commun. Lett.},
  vol.~9, no.~2, pp. 210--214, Feb. 2020.

\bibitem{Shtaiwi2021}
{E. Shtaiwi et al.}, ``Channel estimation approach for {RIS} assisted {MIMO}
  systems,'' \emph{IEEE Trans. on Cogn. Commun. Netw.}, vol.~7, no.~2, pp.
  452--465, Jun. 2021.

\bibitem{Zheng2024}
{P. Zheng et al.}, ``{JrCUP}: Joint {RIS} calibration and user positioning for
  {6G} wireless systems,'' \emph{IEEE Trans. Wireless Commun.}, vol.~23, no.~6,
  pp. 6683--6698, Jun. 2024.

\bibitem{Mylonopoulos2022}
G.~Mylonopoulos, C.~D’Andrea, and S.~Buzzi, ``Active reconfigurable
  intelligent surfaces for user localization in {mmWave MIMO} systems,'' in
  \emph{Proc. IEEE Workshop Signal Process. Adv. Wireless Commun. (SPAWC)},
  Jul. 2022, pp. 1--5.

\bibitem{Zhang2023}
{Z. Zhang et al.}, ``Active {RIS} vs. passive {RIS}: Which will prevail in
  {6G}?'' \emph{IEEE Trans. Commun.}, vol.~71, no.~3, pp. 1707--1725, 2023.

\bibitem{Peng2022}
Z.~Peng, X.~Liu, C.~Pan, L.~Li, and J.~Wang, ``Multi-pair {D2D} communications
  aided by an active {RIS} over spatially correlated channels with phase
  noise,'' \emph{IEEE Wireless Commun. Lett.}, vol.~11, no.~10, pp. 2090--2094,
  Oct. 2022.

\bibitem{Huang2021}
Z.~Huang, B.~Zheng, and R.~Zhang, ``Transforming fading channel from fast to
  slow: {IRS}-assisted high-mobility communication,'' in \emph{Prof. IEEE Int.
  Conf. Commun. (ICC)}, Jun. 2021, pp. 1--6.

\bibitem{Peng2023}
{Z. Peng et al.}, ``Two-stage channel estimation for {RIS}-aided multiuser
  {mmWave} systems with reduced error propagation and pilot overhead,''
  \emph{IEEE Trans. Signal Process.}, vol.~71, pp. 3607--3622, 2023.

\bibitem{Wan2020}
Z.~Wan, Z.~Gao, and M.-S. Alouini, ``Broadband channel estimation for
  intelligent reflecting surface aided {mmWave} massive {MIMO} systems,'' in
  \emph{Prof. IEEE Int. Conf. Commun. (ICC)}, Jun. 2020, pp. 1--6.

\bibitem{Lu2022}
Y.~Lu, H.~Chen, J.~Talvitie, H.~Wymeersch, and M.~Valkama, ``Joint {RIS}
  calibration and multi-user positioning,'' in \emph{Proc. IEEE Veh. Technol.
  Conf. (VTC-Fall)}, Sep. 2022, pp. 1--6.

\bibitem{Li2020}
S.~Li, B.~Duo, X.~Yuan, Y.-C. Liang, and M.~Di~Renzo, ``Reconfigurable
  intelligent surface assisted {UAV} communication: Joint trajectory design and
  passive beamforming,'' \emph{IEEE Wireless Commun. Lett.}, vol.~9, no.~5, pp.
  716--720, May 2020.

\bibitem{Kolda2009}
T.~G. Kolda and B.~W. Bader, ``Tensor decompositions and applications,''
  \emph{SIAM Rev.}, vol.~51, no.~3, pp. 455--500, Aug. 2009.

\bibitem{Wen2018}
F.~Wen, N.~Garcia, J.~Kulmer, K.~Witrisal, and H.~Wymeersch, ``Tensor
  decomposition based beamspace {ESPRIT} for millimeter wave {MIMO} channel
  estimation,'' in \emph{Prof. IEEE Global Commun. Conf. (GLOBECOM)}, Dec.
  2018, pp. 1--7.

\bibitem{Wen2020}
F.~Wen, H.~C. So, and H.~Wymeersch, ``Tensor decomposition-based beamspace
  {ESPRIT} algorithm for multidimensional harmonic retrieval,'' in \emph{Proc.
  IEEE Int. Conf. Acoust. Speech Signal Process. (ICASSP)}, May 2020, pp.
  4572--4576.

\bibitem{Jiang2001}
T.~Jiang, N.~Sidiropoulos, and J.~ten Berge, ``Almost-sure identifiability of
  multidimensional harmonic retrieval,'' \emph{IEEE Trans. Signal Process.},
  vol.~49, no.~9, pp. 1849--1859, Sep. 2001.

\bibitem{Wax1985}
M.~Wax and T.~Kailath, ``Detection of signals by information theoretic
  criteria,'' \emph{IEEE Trans. Acoust. Speech Signal Process.}, vol.~33,
  no.~2, pp. 387--392, Apr. 1985.

\bibitem{Zhou2017}
Z.~Zhou, J.~Fang, L.~Yang, H.~Li, Z.~Chen, and R.~S. Blum, ``Low-rank tensor
  decomposition-aided channel estimation for millimeter wave {MIMO-OFDM}
  systems,'' \emph{IEEE J. Sel. Areas Commun.}, vol.~35, no.~7, pp. 1524--1538,
  Jul. 2017.

\bibitem{Sengupta1993}
S.~M. Kay, \emph{Fundamentals of statistical signal processing: Estimation
  theory}.\hskip 1em plus 0.5em minus 0.4em\relax Hoboken, NJ, USA: Prentice
  Hall, 1993.

\bibitem{Huang2022}
Z.~Huang, B.~Zheng, and R.~Zhang, ``Transforming fading channel from fast to
  slow: Intelligent refracting surface aided high-mobility communication,''
  \emph{IEEE Trans. Wireless Commun.}, vol.~21, no.~7, pp. 4989--5003, Jul.
  2022.

\bibitem{Horn2012}
R.~A. Horn and C.~R. Johnson, \emph{Matrix analysis}.\hskip 1em plus 0.5em
  minus 0.4em\relax Cambridge, U.K.: Cambridge Univ. Press, 2012.

\bibitem{Vervliet2016}
\BIBentryALTinterwordspacing
N.~Vervliet, O.~Debals, L.~Sorber, M.~Van~Barel, and L.~De~Lathauwer,
  ``Tensorlab 3.0,'' Mar. 2016. [Online]. Available:
  \url{https://www.tensorlab.net}
\BIBentrySTDinterwordspacing

\bibitem{Boumal2014}
\BIBentryALTinterwordspacing
N.~Boumal, B.~Mishra, P.-A. Absil, and R.~Sepulchre, ``{M}anopt, a {M}atlab
  toolbox for optimization on manifolds,'' \emph{J. Mach. Learn. Res.},
  vol.~15, no.~42, pp. 1455--1459, Apr. 2014. [Online]. Available:
  \url{https://www.manopt.org}
\BIBentrySTDinterwordspacing

\bibitem{Chen2024}
A.~Chen, L.~Chen, Y.~Chen, N.~Zhao, and C.~You, ``Near-field positioning and
  attitude sensing based on electromagnetic propagation modeling,'' \emph{IEEE
  J. Sel. Areas Commun.}, vol.~42, no.~9, pp. 2179--2195, Sep. 2024.

\bibitem{Zhang2024c}
{Z. Zhang el al.}, ``{Ziv–Zakai} bound for {2D-DOAs} estimation,'' \emph{IEEE
  Trans. Signal Process.}, vol.~72, pp. 2483--2497, 2024.

\bibitem{Zhang2023a}
Z.~Zhang, Z.~Shi, and Y.~Gu, ``{Ziv-Zakai} bound for {DOAs} estimation,''
  \emph{IEEE Trans. Signal Process.}, vol.~71, pp. 136--149, 2023.

\end{thebibliography}

\end{document}